\documentclass[a4paper,fleqn,usenatbib]{mnras}
\usepackage{newtxtext,newtxmath}
\usepackage[T1]{fontenc}
\usepackage{ae,aecompl}
\usepackage{graphicx,psfig}	
\usepackage{epsfig}	
\usepackage{amsmath}	
\usepackage{amssymb}	
\usepackage{graphics}
\usepackage{longtable}
\usepackage[font=small,labelfont=bf,margin=\parindent,tableposition=top]{caption}
\usepackage[usenames,dvipsnames,svgnames,table]{xcolor}

\usepackage[switch]{lineno} 
\usepackage{lipsum} 
\usepackage{hyperref}
 \definecolor{arancio}{rgb}{1,0.5,0}
 \definecolor{viola}{rgb}{0.7,0,1}
 \definecolor{verde}{rgb}{0.2,0.7,0.7}
\newcommand\pmn{\ifmmode {\rm PMN~J0948}+0022 \else PMN~J0948$+$0022\fi} 
\newcommand\sbs{\ifmmode {\rm SBS~0846}+513 \else SBS~0846$+$513\fi} 
\newcommand\pks{\ifmmode {\rm PKS~1502}+036 \else PKS~1502$+$036\fi} 
\newcommand\kms{\ifmmode {\rm km\ s}^{-1} \else km s$^{-1}$\fi} 
\newcommand\FWHM{\ifmmode {\rm FWHM} \else ${\rm FWHM}$\fi}
\newcommand\Lsun{\ifmmode L_{\odot} \else $L_{\odot}$\fi} 
\newcommand\Hbeta{\ifmmode {\rm H}\beta  \else H$\beta$\fi} 

\def \ATel {ATel} 
\def \apj {ApJ}
\def \apjl {ApJL}

\def \aap {A\&A}

\def \mnras {MNRAS}

\title[$\gamma$-$\gamma$ absorption in NLS1s]{
Prospects 
for gamma-ray observations of narrow-line Seyfert 1 galaxies with the Cherenkov Telescope Array II. \\  
Gamma-gamma absorption in the broad-line region radiation fields}

\author[P. Romano et al.]{
P.\ Romano,$^{1}$\thanks{E-mail: patrizia.romano@inaf.it}$^{,2}$\thanks{Visiting Astronomer at LUTH.}
M.\ B\"ottcher,$^{3}$
L.\ Foschini,$^{1}$
C.\ Boisson,$^{2}$ 
S.\ Vercellone,$^{1}$
\newauthor  
and M.\ Landoni$^{1}$ \\
$^{1}$INAF, Osservatorio Astronomico di Brera, Via E.\ Bianchi 46, I-23807, Merate, Italy\\
$^{2}$LUTH, Observatoire de Paris, CNRS, Universit\'e  Paris Diderot, PSL Research University Paris, 5 place Jules Janssen, \\
F-92195 Meudon, France \\ 
$^{3}$Centre for Space Research, North-West University, Potchefstroom, 2531, South Africa\\ 
}

\date{Accepted 2020 February 25. Received 2020 February 18; in original form 2020 January 14}

\pubyear{2020}

\begin{document}
\label{firstpage}
\pagerange{\pageref{firstpage}--\pageref{lastpage}}
\maketitle

\begin{abstract}
Gamma-ray emitting narrow-line Seyfert 1 ($\gamma$-NLS1) galaxies possibly  
harbour relatively low-mass black holes (10$^6$--10$^8$\,M$_{\sun}$) accreting 
close to the Eddington limit, and share many 
characteristics with their sibling sources, flat-spectrum radio quasars.
Although they have been detected in the MeV--GeV band with Fermi-LAT, 
they have never been seen in the very high energy band with current imaging atmospheric Cherenkov telescopes (IACTs). 
Thus, they are key targets for the next-generation IACT, the Cherenkov Telescope Array  (CTA).
In a previous work we selected, by means of extensive simulations, the best candidates 
for a prospective CTA detection (SBS~0846$+$513,  PMN~J0948$+$0022, and PKS~1502$+$036)
taking into  account the effects of both the intrinsic absorption 
(approximated with a cut-off at 30\,GeV), and the extra-galactic background light 
on the propagation of $\gamma$-rays. 
In this  work we simulate the spectra of these three sources by adopting more realistic 
broad-line region (BLR) absorption models. 
In particular, we consider the detailed treatment of $\gamma$-$\gamma$ absorption 
in the radiation fields of the  BLR as a function of the location of the $\gamma$-ray emission region
with parameters inferred from observational constraints. 
We find that, due to the energy range extent and its sensitivity, 
CTA is particularly well suited to locate the $\gamma$-ray emitting region in $\gamma$-NLS1. 
In particular CTA will be able not only to distinguish whether the $\gamma$-ray emitting region is located 
inside or outside the BLR, but also where inside the BLR it may be. 
\end{abstract}



 	 \section{Introduction}  \label{romano_nls1_2:intro}

\setcounter{table}{0} 
 \begin{table*} 
\begin{center}
\small	
 \tabcolsep 5pt  
 \caption{Sample of $\gamma$-NLS1, and the parameters of their $\gamma$-ray spectra in several flux states, 
optically-derived luminosities, adopted for the simulations. 
}
  \label{romano_nls1_2:table:sample} 
  \begin{tabular}{lc rr r  cccc cc} 
 \hline 

 \noalign{\smallskip} 
    Source Name                   &   State & RA     & Decl        & $z$    &   $K_{0}$                       &  $E_{0}$  &  $\Gamma$  & Ref. &  \hspace{+6pt}L$_{\rm H\beta}$$^a$    & L$_{\rm disc}$$^b$      \\
                                           &           &  (deg) & (deg)	 &          & (ph\,cm$^{-2}$\,MeV$^{-1}$\,s$^{-1}$)    & (MeV)   &         \\
 \noalign{\smallskip} 
 \hline 
 \noalign{\smallskip} 
SBS~0846$+$513       &   High           &132.51    &51.14  & 0.585   &1.08$\times10^{-10}$	&300   	& 2.10   &  1  &1.32   &3.94   \\  
PMN~J0948$+$0022  &   High           &147.24    &0.37    & 0.585  &9.60$\times10^{-10}$	&300	 &2.55   &  2  &3.73    &11.8 \\
                                  &   ``Flare''      & -	    & -	 &   -        &2.88$\times10^{-9}$	&300	 &2.55   &  3  & -        & - \\
PKS~1502$+$036	  &  High            &226.26    &3.44    & 0.408   &1.4$\times10^{-9}$	&250 	 &2.54   &  4  &0.41    &1.12 \\
\noalign{\smallskip} 
  \noalign{\smallskip}  
  \hline
\end{tabular}
\end{center}
\begin{list}{}{} 
  \item {Redshift are drawn from NED.  
Luminosities  estimated from optical (SDSS) data, as drawn from \citet[][]{Foschini2015:fsrl_nls1}. 
  \item {$^a$}{H$\beta$ luminosities in units of $10^{42}$ erg s$^{-1}$.}
  \item {$^b$}{Disc luminosities in units of $10^{44}$ erg s$^{-1}$.} 
}
\end{list}   
\begin{list}{\it References.}{} 
\item For the power-law models we adopted:  
(1) \citet[][]{Paliya2016:0846}; 
(2) \citet[][]{Foschini2011:J0948}; 
(3) See \citet[][]{Romano2018:nls1_cta}: flaring state, assumed a factor of 3 brighter than the high state; 
(4) \citet[][]{Dammando2016:1502}. 
\end{list}   
\end{table*} 

Narrow-Line Seyfert 1 galaxies (NLS1s) are a subclass of active galactic nuclei (AGN) whose optical properties, 
with narrow permitted emission lines (\Hbeta{} \FWHM $<$ 2000 $\kms$, \citealt[][]{Goodrich1989:nls1def}),  
weak forbidden [\ion{O}{III}] lines ([\ion{O}{III}]\,$\lambda 5007/ \Hbeta < 3$), 
and strong iron emission lines (high \ion{Fe}{II}/\Hbeta, \citealt[][]{OsterbrockP1985:nls1def}),
set them apart from the more general population of Seyfert 1 galaxies (broad-line Seyfert 1s, BLS1s). 
These characteristics \citep[e.g.][]{Peterson2004} are generally explained in terms of lower masses ($10^6$--$10^8$ M$_{\sun}$) 
of the central black hole when compared to BLS1s with similar luminosities, and higher accretion rates, 
close to the Eddington limit
(but see, also, \citealt{Viswanath2019:NLS1masses} and references therein). 
Recent evidence has been found that a small fraction 
(4--7\,\%, \citealt[][]{Komossa2006:rlnl1q,Cracco2016}) 
of NLS1s are radio loud and show a flat radio spectrum 
(\citealt[][]{Oshlack2001:pks2004-447,Zhou2003:0948,Yuan2008}; 
see also, \citealt[][]{Lahteenmaki2017}). 
Further evidence of a hard component in some of their X-ray spectra  
and spectral variability in the hard X-ray 
was also found \citep[][]{Foschini2009:Adv}. 

Following the first detection of a NLS1 in $\gamma$-rays (E $> 100$\,MeV) by 
{\it Fermi}-LAT \citep[PMN~J0948$+$0022,][]{Abdo2009:J0948discov,Foschini2010:J0948,Abdo2009:J0949mw}
a new subclass of NLS1s was defined, the $\gamma$-ray emitting NLS1 ($\gamma$-NLS1) galaxies,
now consisting of about 20 objects, as sources whose 
overall observational properties are strongly reminiscent of those of jetted sources 
\citep[see, e.g.][]{Foschini2012:review,Foschini2015:fsrl_nls1,Dammando2016:jets_nls1}. 
Currently, however, no detection has been obtained in the 
very high energy (VHE, $E>50$\,GeV) regime 
(Whipple upper limit (UL) on 1H~0323$+$342, \citealt{Falcone2004}; 
VERITAS UL on PMN~J0948$+$0022, \citealt{Dammando2015:J0948}; 
H.E.S.S. UL on PKS 2004$-$447, \citealt{2014A&A...564A...9H}). 

In \citet[][]{Romano2018:nls1_cta} (Paper I)
we considered the prospects for observations of 
$\gamma$-NLS1 as a class of sources to investigate with the Cherenkov Telescope Array (CTA)  
since the detection in the VHE regime would provide important clues on the location of the 
$\gamma$-ray emitting region.
The CTA, as the next-generation ground-based  $\gamma$-ray observatory, 
will boast a wide energy range (20\,GeV to 300\,TeV) which will be achieved 
by including 3 classes of telescopes with different sizes, i.e., 
the large-sized telescopes (LSTs, diameter D$\sim23$\,m),   
the medium-sized telescopes (MSTs, D$\sim12$\,m)  
and the small-sized telescopes (SSTs, primary mirror D$\sim4$\,m).  
CTA will also provide all-sky coverage, by consisting of two separate arrays on two sites, one 
in each hemisphere. 
The current CTA setup \citep[][]{2017AIPC.1792b0014H,2017Msngr.168...21H} 
includes a Northern site at the Observatorio del Roque de los Muchachos on the island of La Palma (Spain) 
where 4 LSTs and 15 MSTs, covering an area of $\sim1$\,km${^2}$, will be located,
and a Southern site at the European Southern Observatory's (ESO's) Paranal Observatory 
in the Atacama Desert (Chile), that will  cover an area of about 4\,km$^{2}$, where 
4 LSTs, 25 MSTs, and 70 SSTs will be located. 

The  extensive set of simulations of all currently known $\gamma$-ray emitters 
identified as NLS1s (20 sources)  reported in \citet[][]{Romano2018:nls1_cta}  
took into  account the effect of both the extra-galactic background light on the
propagation of $\gamma$-rays and intrinsic absorption components. 
These latter components, mainly 
due to the currently unconstrained location of the $\gamma$-ray emitting region, 
were approximated analytically with a cut-off at 30\,GeV ($\propto e^{-E/E_{\rm cut}}$, $E_{\rm cut}=30$\,GeV).
In this work we consider the only three sources that were 
deemed good candidates for a prospective CTA detection 
in \citet[][]{Romano2018:nls1_cta}, \sbs, \pmn, and \pks,
and simulate their spectra by adopting more realistic broad-line region (BLR) absorption models. 
In particular, we shall consider the detailed treatment of $\gamma$-$\gamma$ absorption 
in the radiation fields of the broad-line region (BLR) of these NLS1s 
as a function of the location of the $\gamma$-ray emission region as 
proposed by \citet[][]{Bottcher2016:blr}. 

In Sect.~\ref{romano_nls1_2:sample} we detail our sample; in 
Sect.~\ref{romano_nls1_2:sim_blr_abs} we describe the BLR internal absorption modelling, in 
Sect.~\ref{romano_nls1_2:simulations} we describe our simulation setup. 
In Sect.~\ref{romano_nls1_2:results} we present our results 
and in Sect.~\ref{romano_nls1_2:discussion}  discuss their implications.

 	 \section{Data Sample}  \label{romano_nls1_2:sample}

Of the 20-source sample described in \citet[][Table~1]{Romano2018:nls1_cta}, 
as expected due to the faintness of $\gamma$-NLS1s, 
we only consider \sbs,  \pmn, and \pks, 
which were deemed good candidates for a prospective CTA detection and, 
hence, should afford the chance to extract  meaningful spectra. 
For those 3 sources different activity states were defined/studied in Paper I 
whose properties are recalled below.
The properties of these sources are found in Table~\ref{romano_nls1_2:table:sample}, and include 
coordinates (Equatorial, J2000, Cols.~3, 4), redshift (Col.~5), 
the spectral parameters for the best fit models to the {\it Fermi} data  adopted for 
each source and flux state (Cols.~6--9),  
where the spectra are described as a power-law (PL)
\begin{equation}
\frac{dN}{dE}=K_0 \left( \frac{E}{E_0} \right) ^{-\Gamma},
\end{equation}
where $K_0$ is the normalisation (in units of ph\,cm$^{-2}$\,s$^{-1}$\,MeV$^{-1}$), 
$E_0$ is the pivot energy (in MeV), 
and  $\Gamma$ is the power-law photon index.
Table~\ref{romano_nls1_2:table:sample} also reports the H$\beta$ and disc 
luminosities 
derived from Table~2 of \citet[][]{Foschini2015:fsrl_nls1}, 
that we adopted as the basis of the modelling of the BLR absorption we present in 
Sect.~\ref{romano_nls1_2:sim_blr_abs}. 
In the following we summarise the models we considered for the three sources. 

For \sbs{} we considered only the high-flux state, which was derived from \citet[][]{Paliya2016:0846} 
and was modelled by means of a simple power-law model with photon 
index $2.10\pm0.05$ 
and an integrated $\gamma$-ray flux ($0.1<E<300$\,GeV) 
of $(9.92\pm0.84)\times 10^{-8}$ 
\,ph\,cm$^{-2}$\,s$^{-1}$ (F2 flare, integrated over 120\,days).  

For \pmn{} two flux states were considered. 
The high state ($F_{\rm E>100 MeV}=(1.02\pm0.02)\times10^{-6}$\,ph\,cm$^{-2}$\,s$^{-1}$)  
is described by a simple power-law model with photon index 
$\Gamma=2.55\pm0.02$ \citep[][]{Foschini2011:J0948}.  
We also defined a ``flare'' state as three times brighter than the high state, 
3\,hr long, 
with the same spectral shape 
which could represent a very bright state for this source
and an excellent case to test the specific CTA capabilities in detecting 
bright short transients.

For \pks{} only the high-flux state was considered, as derived from \citet[][1\,day integration]{Dammando2016:1502}
and described by a power-law with a photon index $\Gamma = (2.54 \pm 0.04)$ and
a flux $F_{\rm 0.1<E<300 GeV} = (93 \pm 19) \times 10^{-8}$\,ph\,cm$^{-2}$\,s$^{-1}$.

\begin{figure*} 
\vspace{-7.8truecm}
\centerline{
 \hspace{+0.5truecm} 
                 \includegraphics*[angle=0,width=20.5cm]{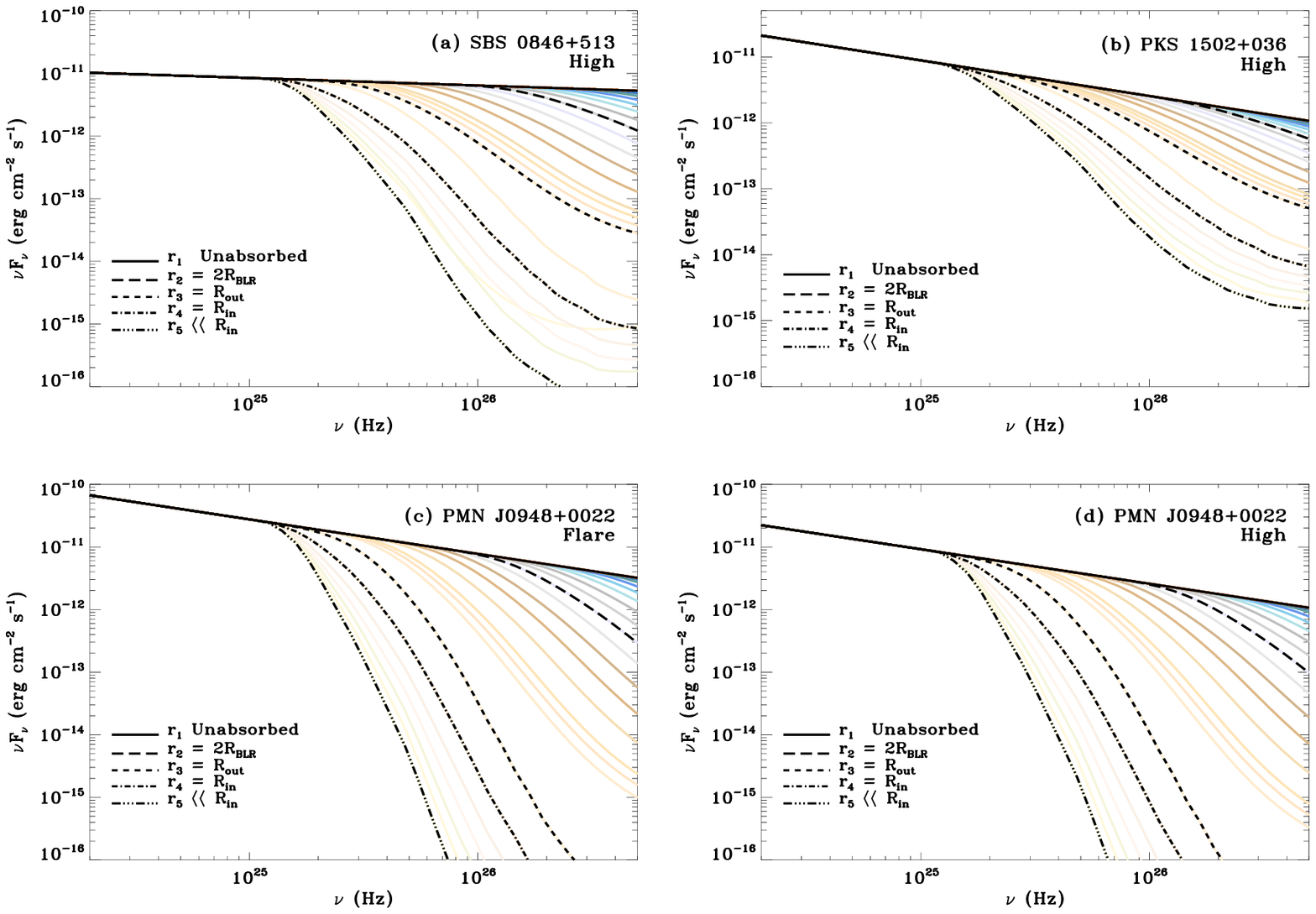}
\vspace{-8.5truecm}
}
\caption[]{Input models including only the BLR contribution to absorption 
(no EBL is taken into account). 
         The grid is calculated for a range of distances, 
from $R_{\rm em} \ll R_{\rm in}$ to $R_{\rm em} \gg R_{\rm out}$. The models in black are 
5 exemplary cases, from least absorbed to most absorbed:  
$r_1 \gg R_{\rm BLR}$, $r_2 =2\, R_{\rm BLR}$, $r_3 =R_{\rm out}$,  $r_4 =R_{\rm in}$, $r_5 \ll R_{\rm in}$,  
while the coloured lines represent models with intermediate radii. 
} 
\label{romano_nls1_2:fig:input_models} 
\end{figure*} 

 	 \section{BLR internal absorption modelling} \label{romano_nls1_2:sim_blr_abs}

\setcounter{table}{1} 
 \begin{table} 
\begin{center}
\small	
 \tabcolsep 2pt  
 \caption{BLR properties adopted for the $\gamma$-$\gamma$ absorption grids 
(Sect~\ref{romano_nls1_2:sim_blr_abs}).}
  \label{romano_nls1_2:table:blr_props} 
  \begin{tabular}{l c c c c c c c} 
 \hline 

 \noalign{\smallskip} 
    Source Name            &   $L_{\rm BLR}^a$        & $R_{\rm BLR}^b$     & $u_{\rm BLR}^c$      & $r_{1}^b$ 	          & $R_{\rm in}^b$   & $R_{\rm out}^b$ & $r_{5}^b$ \\   
                                   &    (erg s$^{-1}$)     &    (cm)       &   (erg cm$^{-3}$)     &   (cm) &   (cm)    &   (cm)           &   (cm)  \\
 \noalign{\smallskip} 
 \hline 
 \noalign{\smallskip} 
SBS~0846$+$513              & 2.8      & 1.87    & 2.12$\times10^{-3}$   & 203 &  1.69  & 2.06 &  0.19 \\   
PMN~J0948$+$0022         & 7.91    & 3.26    & 1.98$\times10^{-3}$   & 351 & 2.93  & 3.58   &  0.33 \\   
PKS~1502$+$036	         & 0.869    & 1.00    & 2.29$\times10^{-3}$   & 108 & 0.90  & 1.10  &  0.10 \\   
\noalign{\smallskip} 
  \hline
\end{tabular}
\end{center}
\begin{list}{}{}
\item {$^a$}{Luminosities in units of $\times10^{43}$ erg s$^{-1}$.}
\item {$^b$}{Radii in units of $\times10^{17}$ cm.}
\item {$^c$}{$u$ in units of $\times10^{-3}$ erg cm$^{-3}$. }
\end{list}   
\end{table} 

For each source and each flux state, we produced a grid of models accounting for 
$\gamma$-$\gamma$ absorption in the BLR as a function of the location of the 
$\gamma$-ray emission region. We followed the methods described in 
\citet[][]{Bottcher2016:blr}, 
but with fixed BLR radius and luminosity derived from observational constraints
as detailed below. 

In 
\citet[][]{Bottcher2016:blr}, the BLR is represented as a spherical, homogeneous shell 
with inner and outer boundaries $R_{\rm in}$ and $R_{\rm out}$, respectively, and the 
$\gamma$-ray emission region is located at a distance $R_{\rm em}$ from the central 
supermassive black hole of the AGN. 
For this work, we assumed that 
the BLR extends from $R_{\rm in} = 0.9 \,R_{\rm BLR}$ to $R_{\rm out} = 1.1\,R_{\rm BLR}$. 
For a given (measured) accretion-disk luminosity, $L_{\rm disk} = 10^{45} \, L_{\rm disc, 45}$~erg~s$^{-1}$, 
the radius of the BLR is estimated as \citep[][]{Bentz2013}, 
     \begin{equation}
       \label{romano_nls1_2:eq:RBLR}
      R_{\rm BLR} = 3 \times 10^{17} L_{\rm disc,45}^{1/2}       {\rm \,\,\,\, [cm]. }
      \end{equation}
The BLR luminosity $L_{\rm BLR}$ is calculated by using the measured luminosity 
of the \Hbeta{} emission line according to the relationship 
      \begin{equation}
       \label{romano_nls1_2:eq:LBLR}
      L_{\rm BLR} = 21.2\,L_{\rm H\beta}        {\rm \,\,\,\, [erg \, s^{-1}]}.   
      \end{equation}
The model BLR spectrum used for the BLR $\gamma$-$\gamma$ absorption calculation includes the 21 
strongest emission lines with relative luminosities taken from the quasar spectrum template by 
\citet[][]{Francis1991}. For the purpose of our $\gamma$-$\gamma$ opacity estimates, the 
stratification of the BLR (with different emission lines originating at different radii within
the BLR) is not taken into account. For a detailed treatment of the effects of such stratification
on the BLR $\gamma$-$\gamma$ opacity, see the Appendix of \cite{Finke2016}.  
The BLR radiation energy density at $R << R_{\rm in}$ is derived from the BLR radius and luminosity
(Eq.~\ref{romano_nls1_2:eq:RBLR} and Eq.~\ref{romano_nls1_2:eq:LBLR}), 
      \begin{equation}
       \label{romano_nls1_2:eq:UBLR}
      u_{\rm BLR}  = \frac{L_{\rm BLR} }{ 4 \pi R_{\rm BLR}^2 c }  {\rm \,\,\,\, [erg \,cm}^{-3}]. 
      \end{equation}

A summary of the  parameters for the BLR of each source is reported in 
Table~\ref{romano_nls1_2:table:blr_props}. 
Grids of models are calculated for a range of distances, 
from $R_{\rm em} \ll R_{\rm in}$ to $R_{\rm em} \gg R_{\rm out}$. 
In Fig.~\ref{romano_nls1_2:fig:input_models} we display the grid of models for each source. 
Among those, we selected 5 exemplary cases, from least absorbed to most absorbed:  
\begin{enumerate}
\item $r_1 \gg R_{\rm BLR}$, 
\item $r_2 =2\, R_{\rm BLR}$,
\item $r_3 =R_{\rm out}$,  
\item $r_4 =R_{\rm in}$, 
\item $r_5 \ll R_{\rm in}$.
\end{enumerate}

\begin{figure*} 
\vspace{-7.8truecm}
\centerline{
 \hspace{+0.2truecm} 
                 \includegraphics*[angle=0,width=20.5cm]{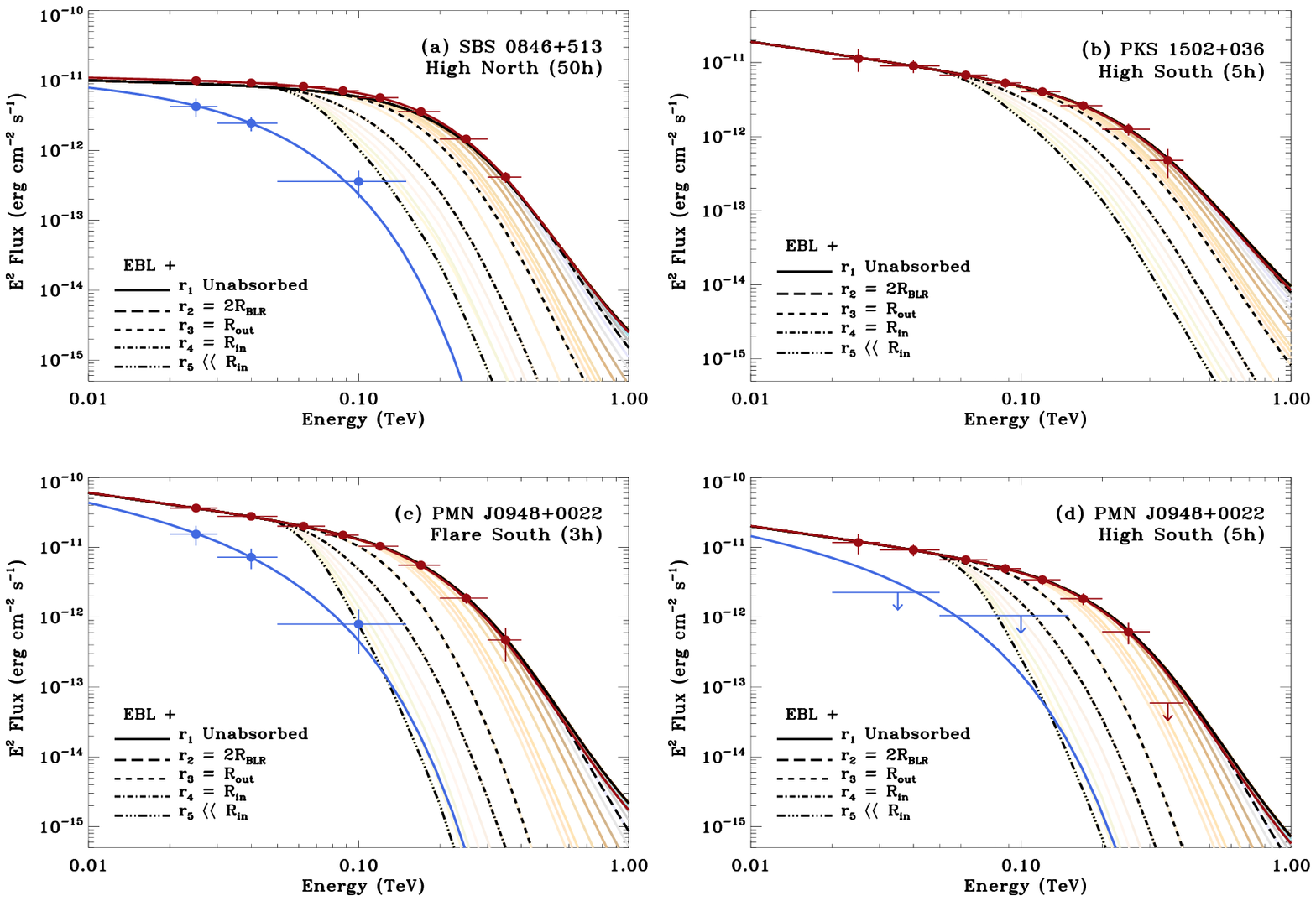}
\vspace{-8.5truecm}
}
\caption[]{Input models including both the BLR  and the EBL contribution to absorption
with the same colour scheme as Fig.~\ref{romano_nls1_2:fig:input_models}. 
Over-plotted are also the input models (blue and red lines) and results (blue and red points) 
of \citet[][Table~1, Figs.~8, 9, 11, and 12]{Romano2018:nls1_cta}: 
the blue lines refer to the power-law model affected by both  EBL and BLR absorption,  
(a cut-off at 30\,GeV, with the exception of the high state of \pks), while 
the red lines refer to models only affected by EBL, that is, assuming that the blue spectra 
extend unbroken above 20--30\,GeV. 
\label{romano_nls1_2:fig:input_models_ebl} 
}
\end{figure*}

Finally, in Fig.~\ref{romano_nls1_2:fig:input_models_ebl} we show the spectral energy 
distributions (SEDs) inputs to our simulations, 
including both the BLR contribution to absorption 
and the attenuation due to the extragalactic background light (EBL, 
calculated by adopting the model of \citealt[][]{Dominguez2011:EBL}).  
Fig.~\ref{romano_nls1_2:fig:input_models_ebl} also shows, as a comparison with previous work, 
the models simulated in \citet[][]{Romano2018:nls1_cta}. 
The blue lines refer to the PL models affected by both EBL and BLR absorption,  
the latter modelled as a cut-off at 30\,GeV, with the exception of the high state of \pks\footnote{In  
\citet[][]{Romano2018:nls1_cta} it was assumed that, as also concluded by \citet[][]{Dammando2016:1502},
due to the 3-week delay observed between the $\gamma$ and radio (15\,GHz) light curve peaks, 
the dissipation region may lie outside the BLR, hence no cut-off  was applied to mimic the BLR absorption.}.  
The red lines refer to models only affected by EBL (no BLR contribution), assuming that the blue spectra 
can extend unbroken above 20--30\,GeV.   
All points are the results of the simulations in \citet[][Figs. 8, 9, 11, and 12]{Romano2018:nls1_cta}.  

%

 	 \section{Simulations}  \label{romano_nls1_2:simulations}
%
 Our simulations were performed with the  analysis package  {\tt ctools}  
\citep[][v.\ 1.5.2]{Gammalib_ctools_2016}\footnote{\href{http://cta.irap.omp.eu/ctools/}{http://cta.irap.omp.eu/ctools/}. }  
and the public CTA instrument response 
files\footnote{\href{https://www.cta-observatory.org/science/cta-performance/}{https://www.cta-observatory.org/science/cta-performance/}. }
(IRF, v.\ prod3b-v1). 
We considered only the instrumental background included in the IRFs 
({\tt CTAIrfBackground}) and no further contaminating astrophysical sources in the 
5\,deg field of view (FOV) we adopted for event extraction. 

Following \citet[][]{Romano2018:nls1_cta}, we simulated  each source from the CTA site 
that provides the largest source elevation. 
The corresponding prod3b-v1 IRFs 
(reported in Table~\ref{romano_nls1_2:table:sims}, Col.~3) were used for the simulations.
We chose ``the average'' IRFs for \pmn{} and \pks, which are visible from both 
hemispheres and simulated the sources from the southern CTA site. 
For SBS 0846+513,  which is only visible  from the Northern hemisphere, 
where the geomagnetic field must be taken into account, 
we chose the azimuth-dependent IRF, corresponding to a pointing toward the 
magnetic North \citep[see, e.g.][]{2017APh....93...76H}. 
We selected the exposures based on our findings in \citet[][]{Romano2018:nls1_cta}, 
and we report them in Column~4 of Table~\ref{romano_nls1_2:table:sims}.
As detailed in Sect.~\ref{romano_nls1_2:sim_blr_abs}, the input spectral model files for {\tt ctools} 
were derived by extrapolating the best-fit {\it Fermi} spectra 
(see Table~\ref{romano_nls1_2:table:sample}) into the CTA energy range. 
We took into account both attenuation due to EBL and internal absorption. 
We selected several energy ranges 
(20--30, 30--50, 50--70, 70--100, 100--140, 140--200, 200--280, 280--400\,GeV) 
and extracted spectral points from the simulated CTA observations.
In each band we first used the task {\tt ctobssim} to create event lists based on our 
input models, and then  {\tt ctlike} to fit a power-law model  
$M_{\rm spectral}(E)=k_0 \left( \frac{E}{E_0} \right) ^{-\Gamma}$ 
(where $k_0$ is the normalisation, 
$E_0$ is the pivot energy, 
and  $\Gamma$ is the power-law photon index) 
by using a maximum likelihood model fitting. 
The normalisation and photon index parameters were free to vary 
while the pivot energy was set to the geometric mean of the boundaries of the energy bin. 
{\tt ctlike} also calculates the test statistic \citep[TS,][]{Cash1979,Mattox1996:Egret} of the 
maximum likelihood model fitting, which we used to assess the goodness of the detection in each band. 
We considered a detection to have a high significance when $TS \geq 25$,    
a low significance when $10 \leq TS < 25$, and not to be detected when $TS <10$. 

We performed sets of $N=100$ statistically independent realisations 
to reduce the impact of variations between individual realisations 
\citep[see, e.g.][]{Gammalib_ctools_2016} and calculated  
the percentage of the detections for $TS >10$ and for $TS >25$, 
the mean TS value and its uncertainty, and the flux mean and its uncertainty. 
When the source was not detected, 
we calculated 95\,\% confidence level upper limits on fluxes from the distribution of the simulated fluxes. 

\setcounter{table}{2} 
 \begin{table}  
 \tabcolsep 3pt  
 \begin{center} 
 \caption{ 
Setup of the ({\tt ctools}) simulations: 
site, IRF, exposure time.
\label{romano_nls1_2:table:sims}} 
  \begin{tabular}{l c cc } 
 \hline 
 \noalign{\smallskip} 
   Source Name (State)                   & \hspace{-3pt}Site$^{\mathrm{a}}$ & IRF                                 & Exp. (h)  \\
\noalign{\smallskip} 
 \hline 
 \noalign{\smallskip} 
 \noalign{\smallskip}
SBS~0846$+$513  (High)     &N & 	{\tt North\_z20\_N\_50h} & 50     \\  
 \noalign{\smallskip}
PMN~J0948$+$0022   (Flare)     &S	&	{\tt South\_z20\_average\_5h}	   & 3   \\  
\noalign{\smallskip}
PMN~J0948$+$0022     (High)   &S	&	{\tt South\_z20\_average\_5h}	   & 5   \\  
 \noalign{\smallskip}

PKS~1502$+$036  (High)     &S	&  	{\tt South\_z20\_average\_5h}         &   5    \\  
 \noalign{\smallskip}
  \hline
  \end{tabular}
\end{center}
\begin{list}{}{} 
   \item[$^{\mathrm{a}}$]{CTA site selected for the simulations: N=North (La Palma, latitude: 28.76 N), S=South (Paranal, latitude: 24.68 S).}
\end{list}   
  \end{table} 

\begin{figure*} 
\vspace{-4.7truecm}
\centerline{
 \hspace{+0.5truecm} 
                 \includegraphics*[angle=0,width=20.5cm]{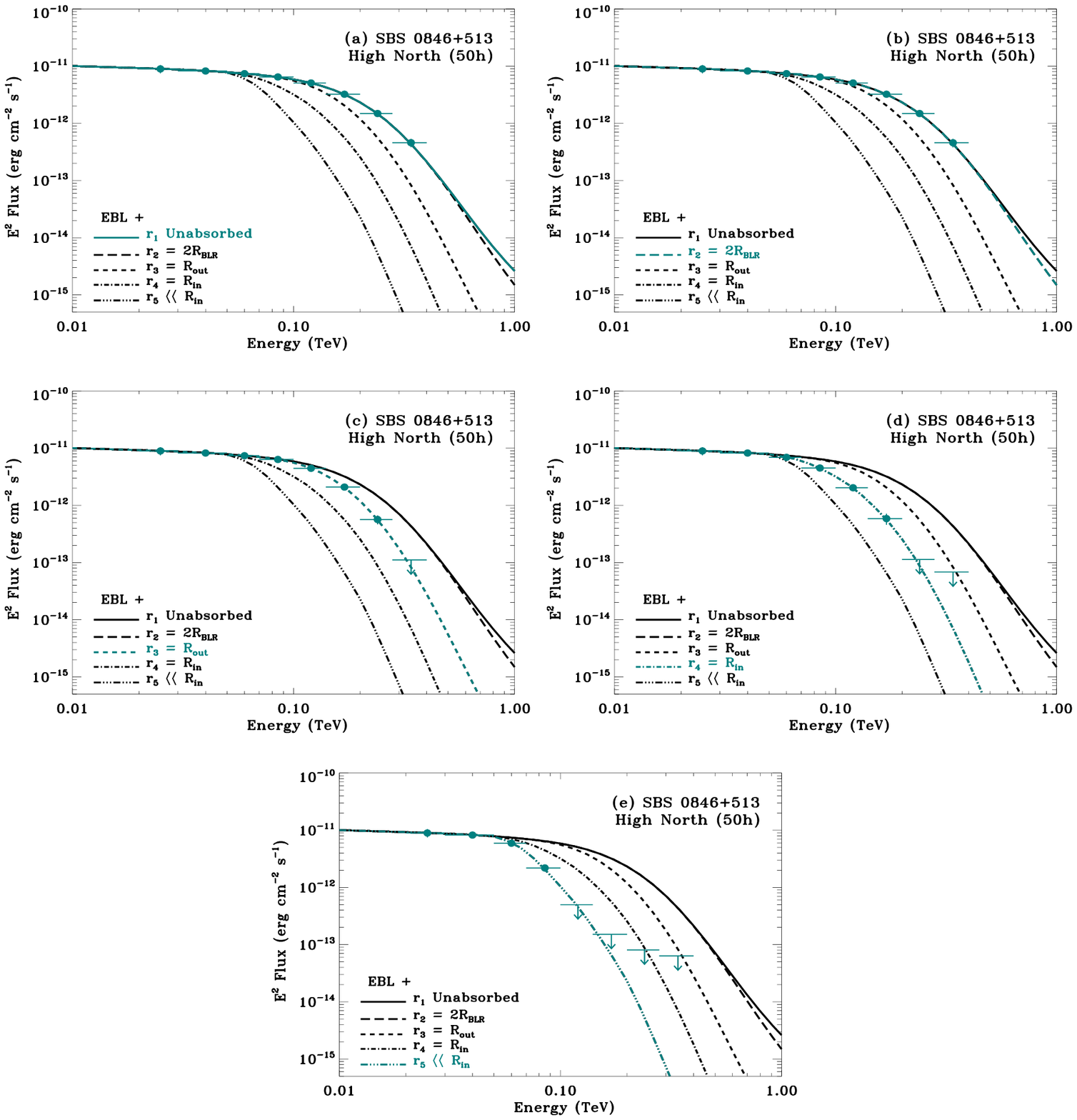} 
\vspace{-5truecm}
}
\caption[]{Input models  for \sbs{}  in the high state, including both the BLR and the EBL contribution to absorption (black). 
The green thick line corresponds to the current input simulated model, while the filled green points represent the results of the simulations. 
The arrows represent the 95\, per cent confidence level upper limits. 
}
\label{romano_nls1_2:fig:res_sbs} 
\end{figure*} 

\begin{figure*} 
\vspace{-4.7truecm}
\centerline{
 \hspace{+0.5truecm} 
                  \includegraphics*[angle=0,width=20.5cm]{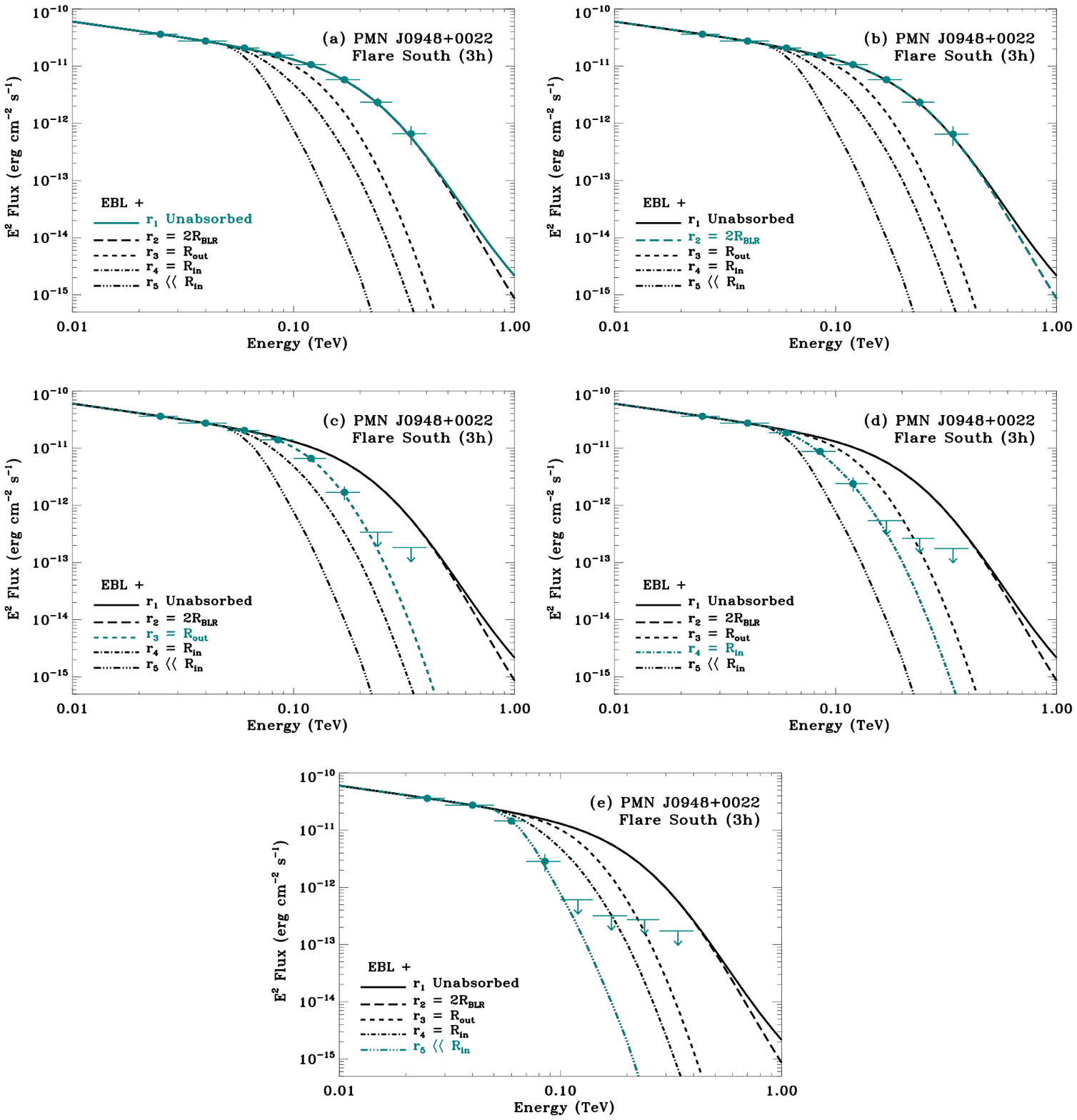} 
\vspace{-5truecm}
}
\caption[]{Same as Fig.~\ref{romano_nls1_2:fig:res_sbs} for \pmn{} in the flaring state.  
}
\label{romano_nls1_2:fig:res_pmnf} 
\end{figure*} 

\begin{figure*} 
\vspace{-4.7truecm}
\centerline{
 \hspace{+0.5truecm} 
                 \includegraphics*[angle=0,width=20.5cm]{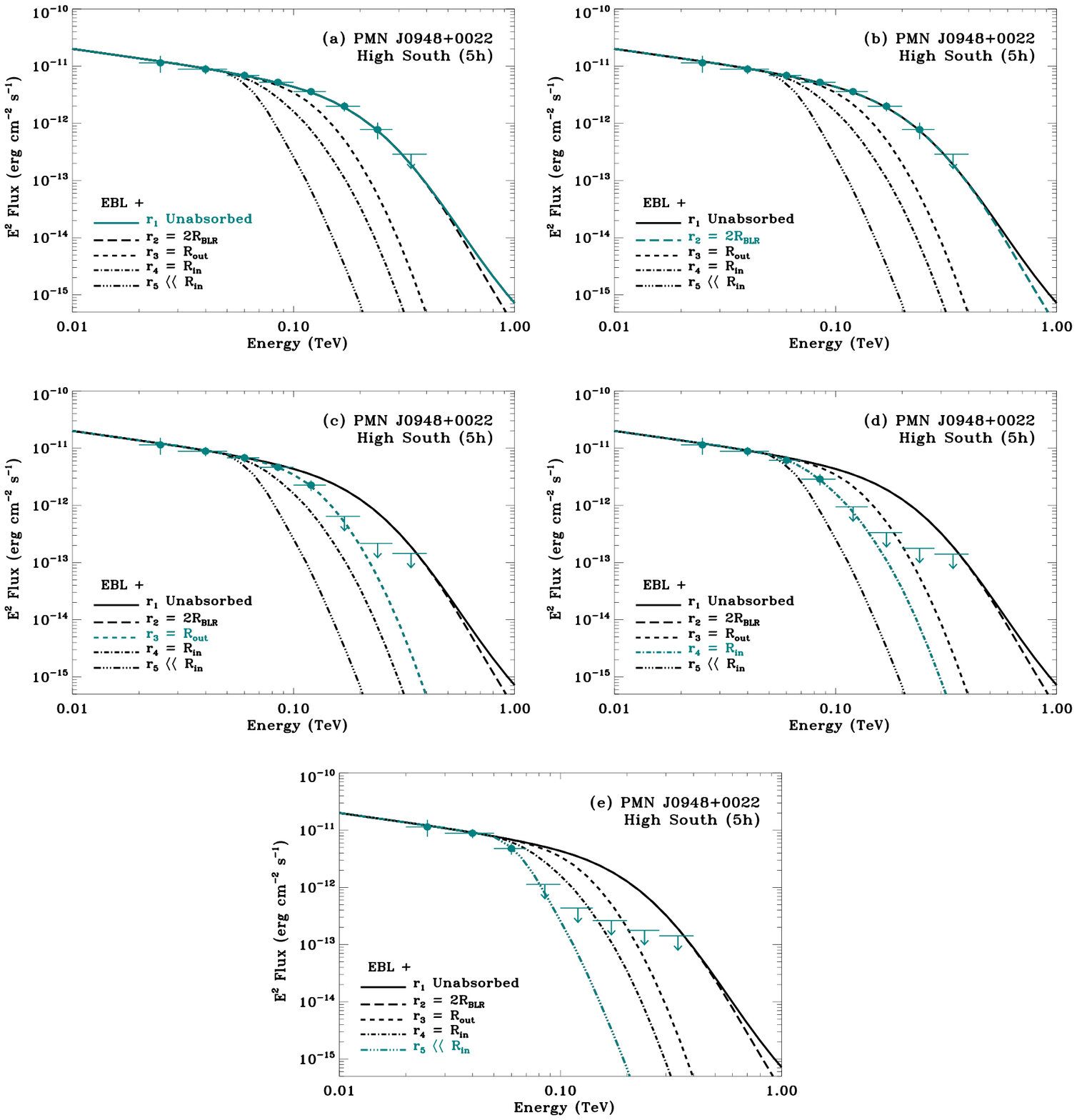} 
\vspace{-5truecm}
}
\caption[]{Same as Fig.~\ref{romano_nls1_2:fig:res_sbs} for \pmn{} in the high state.  
}
\label{romano_nls1_2:fig:res_pmnh} 
\end{figure*} 

\begin{figure*} 
\vspace{-4.7truecm}
\centerline{
 \hspace{+0.5truecm} 
                 \includegraphics*[angle=0,width=20.5cm]{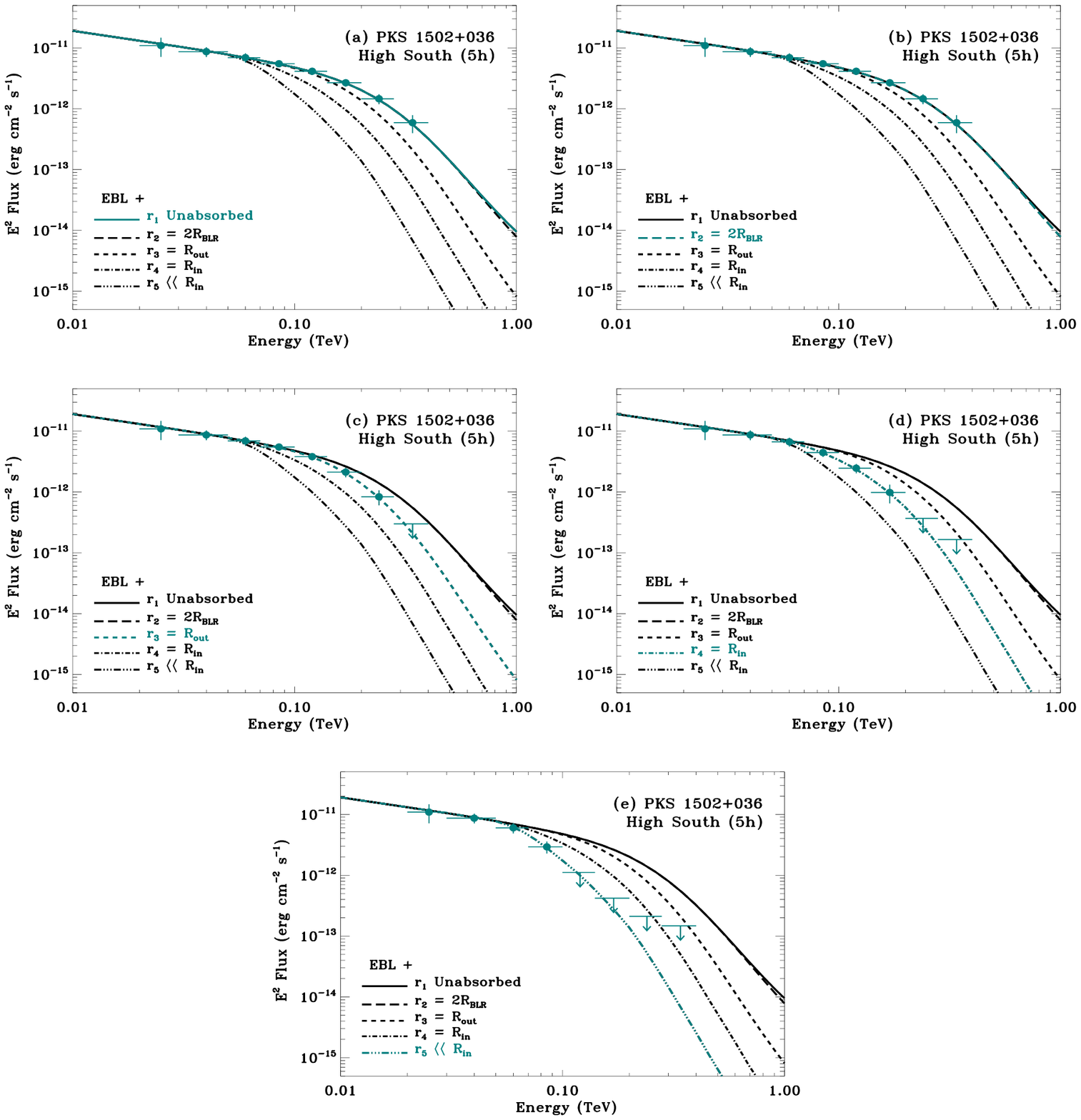} 
\vspace{-5truecm}
}
\caption[]{Same as Fig.~\ref{romano_nls1_2:fig:res_sbs} for \pks{} in the high state.  
}
\label{romano_nls1_2:fig:res_pks} 
\end{figure*} 

 	 \section{Results}  \label{romano_nls1_2:results}

Figure~\ref{romano_nls1_2:fig:res_sbs}--\ref{romano_nls1_2:fig:res_pks} 
shows the results of our simulations. 
For each source, in each panel, 
one of the 5 exemplary models considered is plotted in green
(while the remaining 4 are plotted in black),  
so that the first panel represents the simulations 
for $r_1 \gg R_{\rm out}$ (i.e., the input spectrum unaffected by BLR absorption), 
the second for $r_2 =2\, R_{\rm BLR}$, 
the third for $r_3 =R_{\rm out}$,  
the fourth for $r_4 =R_{\rm in}$, 
the fifth for $r_5 \ll R_{\rm in}$. 
The green points are the simulated fluxes for the current model, and 
the downward pointing arrows are upper limits (95\,\% c.l.). 
Tables~\ref{romano_nls1_2:table:results1}--\ref{romano_nls1_2:table:results4} report the details
of the data points. 

We can see that CTA observations will generally enable us to distinguish models 
with emitting radii within the BLR radius
 ($R_{\rm em} < R_{\rm in}$) 
from those with $R_{\rm em} \gtrsim R_{\rm out}$. 
This is particularly true for \sbs{} in the high state and \pmn{} in flare, and  for \pks{}. 
As expected, for the case of \pmn{} this is more difficult, even in the high state considered
here.

 	 \section{Discussion}  \label{romano_nls1_2:discussion}

%
In \citet[][]{Romano2018:nls1_cta}  we explored the prospects for observations with CTA 
of the largest sample of $\gamma$-NLS1s to date. For our simulations we 
included both the extra-galactic background light in the
propagation of $\gamma$ rays and intrinsic absorption components. 
By adopting a simplified analytical description for the absorption of $\gamma$ rays 
within the source, a cut-off at 30\,GeV ($\propto e^{-E/E_{\rm cut}}$, $E_{\rm cut}=30$\,GeV), 
we could select the best candidates for a prospective CTA detection: 
\sbs{} (high state), \pmn{} (high state), and \pks{} (high and `flare' states, 
see Sect.~\ref{romano_nls1_2:sample}). 

In this paper,  we capitalised on this optimisation of prospective CTA targets and,
motivated by the evidence reported in 
\citet[][]{Berton2016:css_fsrlnls1} that radio-loud NLS1s seem 
to be the low-mass tail of FSRQs, 
we adopted more realistic absorption models as proposed  
for the two FSRQ 3C279 and PKS~1510$-$089 by \citet[][]{Bottcher2016:blr}. 
In that work, this included, within the single-zone leptonic EC-BLR model scenario, 
a detailed treatment of $\gamma$-$\gamma$ absorption in the radiation fields 
of the BLR as a function of the location of the $\gamma$-ray emission region
with parameters inferred from the individual shape of the spectral energy distribution.
For the purpose of this paper, the BLR parameters (radius and luminosity) were kept
fixed based on the observational constraints on the accretion-disk and BLR luminosities
of each of the three NLS1s we examined (Sect.~\ref{romano_nls1_2:sim_blr_abs}).

Figure~\ref{romano_nls1_2:fig:input_models_ebl} compares 
the input models (including both the BLR  and the EBL contribution to absorption) 
of the simplified models of \citet[][]{Romano2018:nls1_cta}, 
in blue (PL $+$EBL $+$BLR absorption) 
and red (PL $+$EBL), and the ones presented here. 
For \sbs{} and \pmn{} the approximation adopted in \citet[][]{Romano2018:nls1_cta}  
was very conservative when compared with the new  models, 
as the shape of the predicted spectrum was quite different and, 
in the case of absorbed models, considerably fainter  for energies below 100\,GeV.  
Also, since the new models diverge sufficiently only above $\sim 50$\,GeV, and  the CTA 
sensitivity\footnote{\href{https://www.cta-observatory.org/science/cta-performance/\#1525680063092-06388df6-d2af}{www.cta-observatory.org/science/cta-performance/}  \#1525680063092-06388df6-d2af.} 
is particularly competitive with respect to {\it Fermi} in that energy range, especially for 
short time-scale phenomena \citep[see][]{ScienceCTA2017,Bulgarelli2015:ICRC,Fioretti2015:ICRC},
CTA is particularly well suited to disentangle these models. 

We also note that our new models are still based on very conservative assumptions
with respect to other models \citep[see][]{Tavecchio2009:3C279},  
first because of the factor of 3 in eq~\ref{romano_nls1_2:eq:RBLR}, 
which differs from, e.g., \citet[][]{Ghisellini2008} but is consistent with 
\citet[][]{Bentz2013}; 
and second because of the assumed BLR geometry, a spherical homogeneous shell, 
which provides the highest optical depth when compared with other geometries, 
for example a ring (see, e.g., Fig.~15 of \citealt[][]{Finke2016}). 
Also, as shown by \citet[][]{Abdo2009:J0948discov}, any reasonable BLR geometry 
does not significantly affect the model results. 

In principle, both the torus and the accretion disc could be considered as 
further sources of absorption. 
However, at energies $E \la 400$\,GeV, where NLS1s might plausibly be detected, 
the torus contribution to absorption is probably irrelevant, as only a detection at $\sim 1\,$TeV
would probe a torus component \citep[see][Fig.~14]{Finke2016}. 
Similarly, the disc is only dominant for $r <10^{16}$\,cm \citep[see][Fig.~15]{Finke2016}. 

In our modelling, we assumed that the curvature in the spectrum is exclusively due to 
$\gamma$-$\gamma$ absorption, but we note that 
there could be additional curvature due to a break or curvature in the e$^-$ spectrum 
or to Klein-Nishina effect. 
Such intrinsic curvature could adversely affect their detectability in CTA observations.  

\begin{figure} 
 \vspace{-0.3truecm}
\begin{flushleft}
                 \epsfig{figure=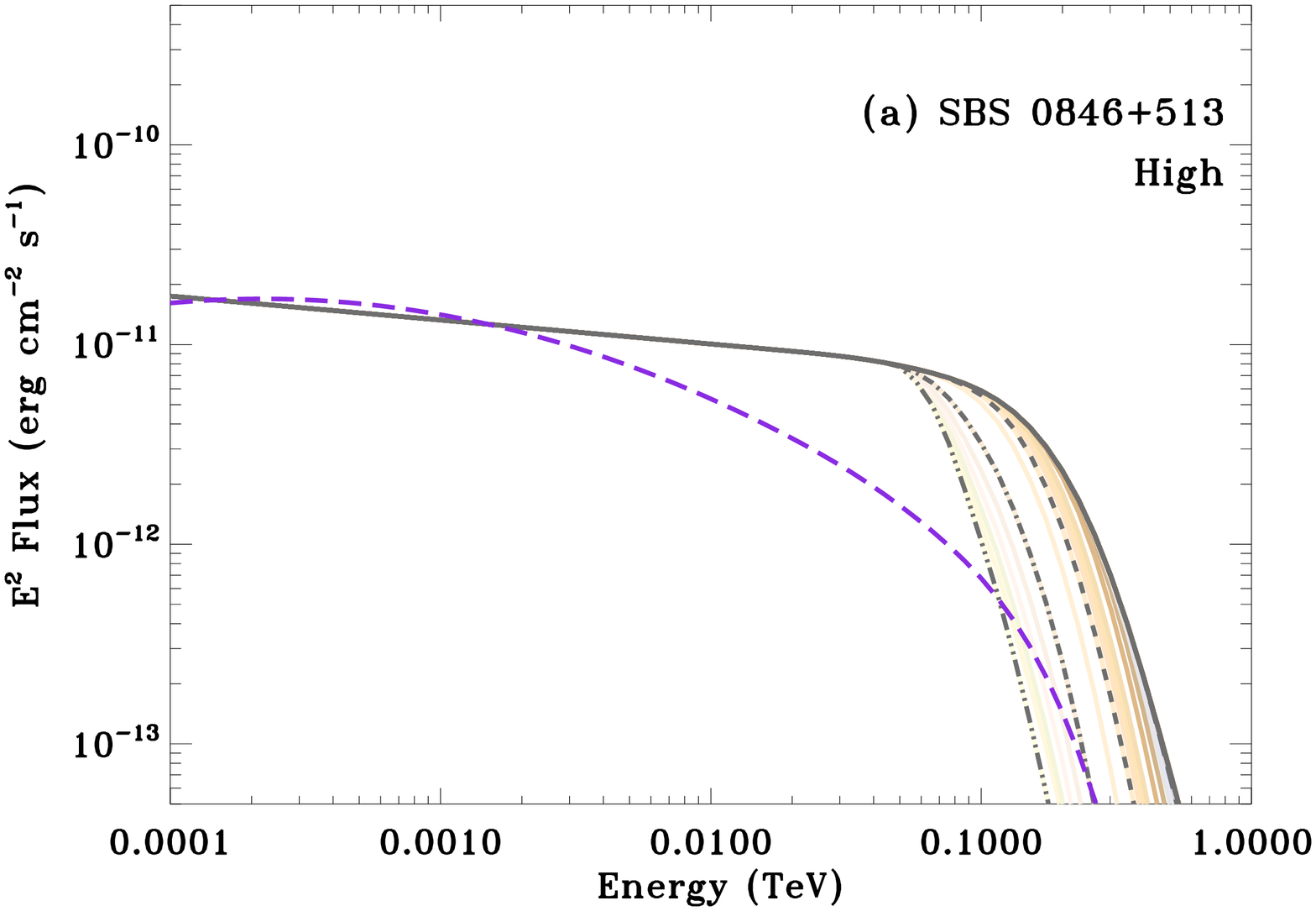,height=8.5cm,angle=90}

\vspace{-0.4truecm} 
                 \epsfig{figure=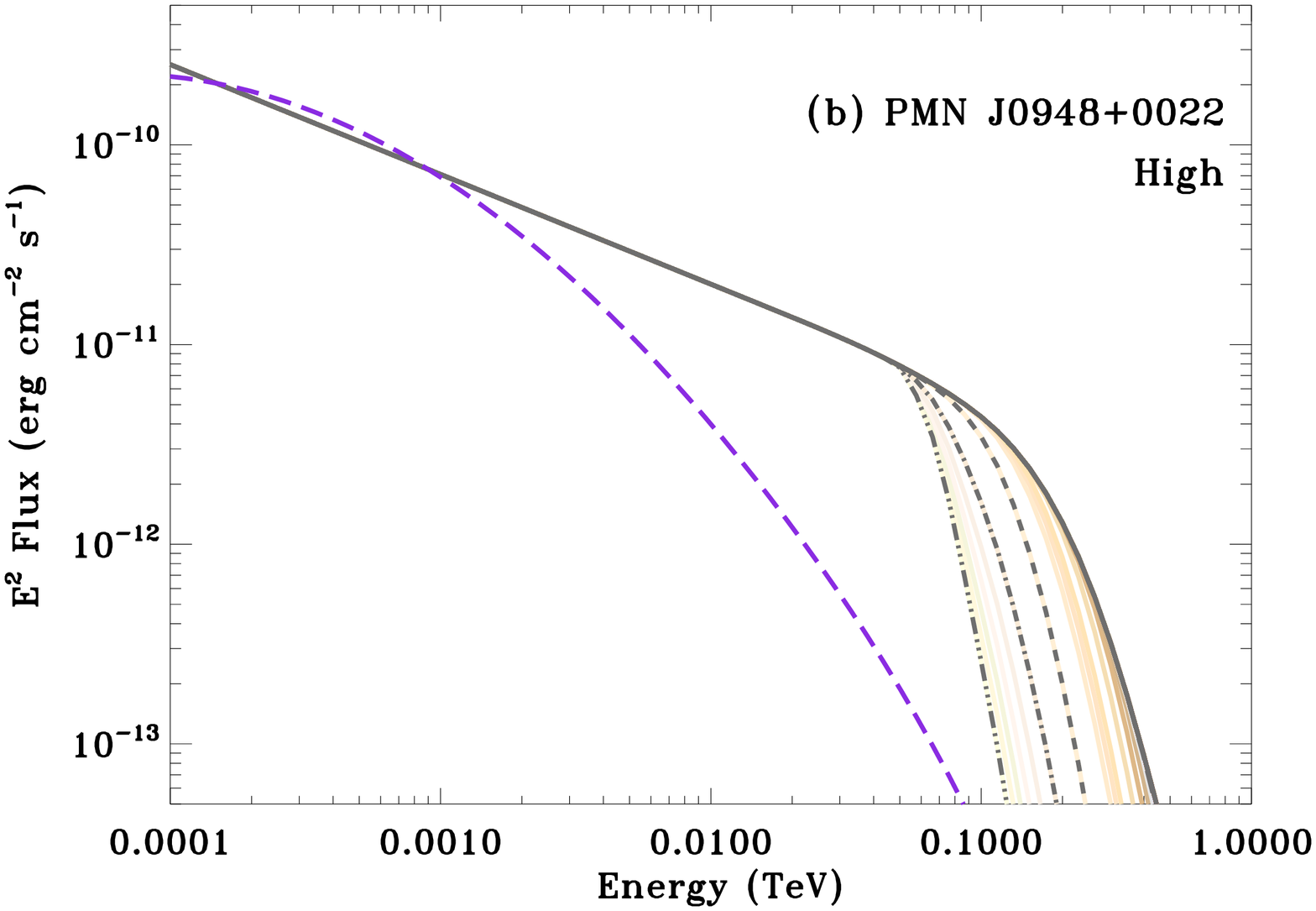 ,height=8.5cm,angle=90}

\vspace{-0.4truecm} 
            \epsfig{figure=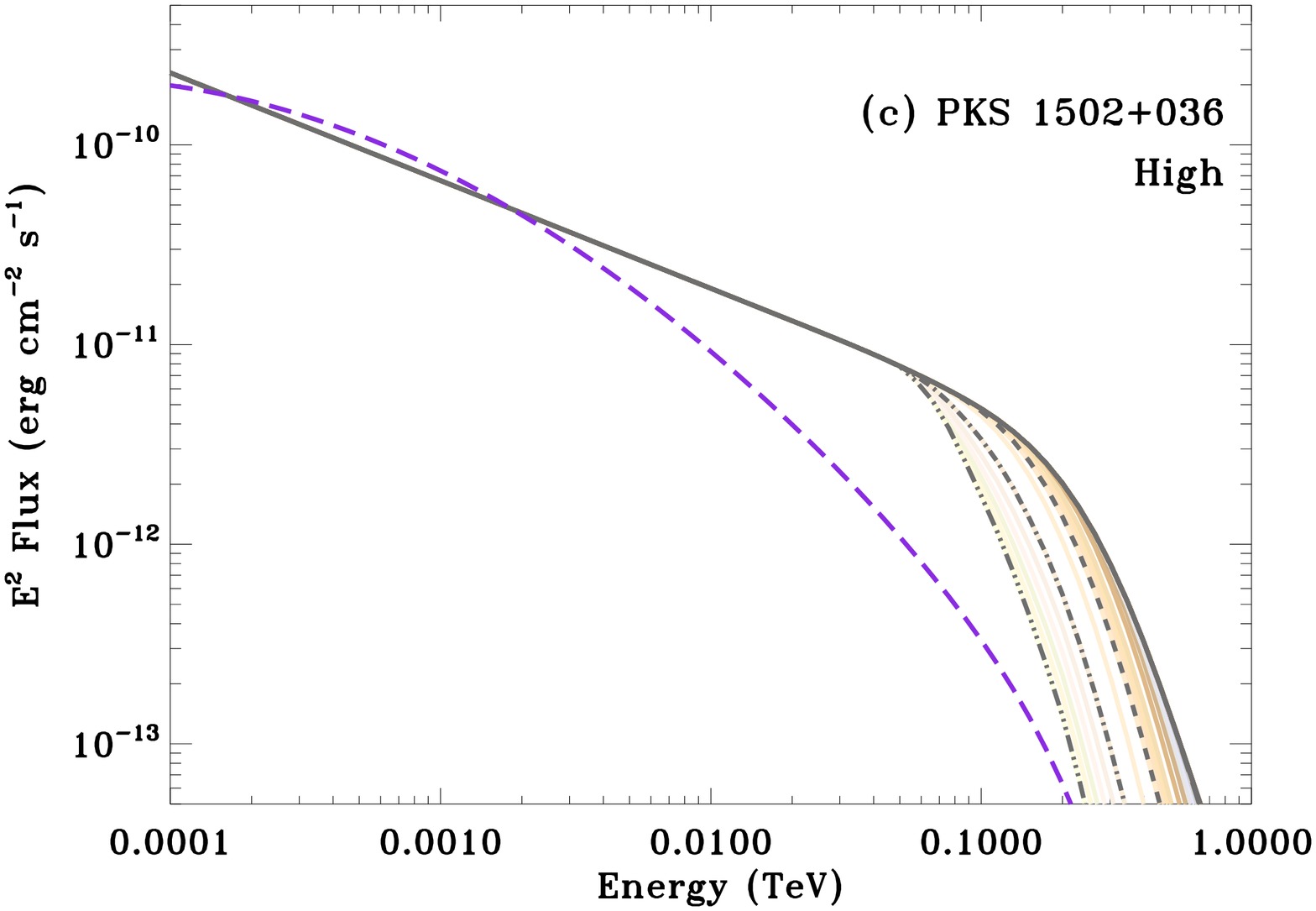 ,height=8.5cm,angle=90 }
 \end{flushleft}

\caption[]{Comparison of the input models for our simulations and the 
extrapolation of the 4FGL log-parabola (violet long-dashed curve) fits scaled at the same Fermi flux. 
}
\label{romano_nls1_2:fig:logpar} 
\end{figure}

In order to address this concern, we can take 
advantage of the extent of the CTA energy range 
to consider a comparison with the extrapolation of a log-parabolic (LP) model fit to the {\it Fermi} data. 
The LP model is described  by 
\begin{equation}
\frac{dN}{dE}=K_0 \left( \frac{E}{E_0} \right) ^{-\alpha -\beta \ln \left(E/E_0\right)},
\end{equation}
where $K_0$ is the normalisation, $E_0$ is the pivot energy,  
$\alpha$ is the local spectral slope at $E = E_0$, and $\beta$ the curvature.  
We drew the LP models for the three sources from the 
Fermi Large Area Telescope Fourth Source Catalog 
\citep[][]{2019arXiv190210045T}, 
and renormalised to the integrated 100\,MeV to 100\,GeV flux 
as extrapolated by adopting the PL models detailed in Table~\ref{romano_nls1_2:table:sample}. 
The renormalisation was obtained by keeping $E_0$, $\alpha$ and $\beta$ fixed. 
Figure~\ref{romano_nls1_2:fig:logpar} shows this comparison for the high states of \sbs, \pmn, and \pks, 
where the LP model is represented by a violet long-dashed curve. 
Even in this case, it will be fairly easy for CTA to distinguish among the models. 

Our simulations, reported in 
Fig.~\ref{romano_nls1_2:fig:res_sbs}--\ref{romano_nls1_2:fig:res_pks},
show that for \sbs{} and \pks{} in the high state and \pmn{} in flare 
it is possible for CTA to distinguish among our  models based on $\gamma$-$\gamma$ 
absorption with emitting radii inside the BLR ($R_{\rm em} < R_{\rm in}$) 
from those with $R_{\rm em} \gtrsim R_{\rm out}$. In the case of 
low-synchrotron-peaked blazars (in particular, FSRQs), leptonic modeling of the 
SEDs typically requires a significant contribution of the BLR to the external 
radiation field for Compton up-scattering to produce the observed {\it Fermi}-LAT 
fluxes and spectra \citep[e.g.,][]{Ghisellini2010,Bottcher2013}. This, along with 
the rapid (sub-hour) variability observed in many FSRQs \citep[e.g.,][]{Ackermann2016,Shukla2018}, 
argues against a location of the $\gamma$-ray emission far beyond the BLR, at $\gg$~pc 
scales in FSRQs. 
At the same time, 
at least for FSRQs with prominent accretion-disk and BLR emission, the same kind of 
BLR absorption constraints evaluated here for $\gamma$-NLS1s, precludes a location of the 
$\gamma$-ray emission region inside the BLR 
\citep[e.g.,][]{Donea2003,Reimer2007,Liu2008,Sitarek2008,Bottcher2016:blr}.
In the case of several observations, especially of FSRQs, this led to the conclusion that the 
most likely location is at $R_{\rm em} \sim R_{\rm out}$ 
\citep[e.g.,][]{Barnacka2014,HESS2019:3C279,HESS2019:PKS0736}, although counter-examples
indicating a location of the $\gamma$-ray emission region at several pc from the central
engine also exist \citep[e.g.][]{Agudo2011}.
Finally, it is also worth noting the instances in which location of $\gamma$-ray emission in the same source changes in different epochs 
\citep[][]{Foschini2008:2155,Foschini2011:FermiSymp,Ghisellini2013:pmn2345,Pacciani2014:blazar_zone,Ahnen2015:pks1441}.  
The similarity of the non-thermal (likely jet-dominated) components of the SEDs of $\gamma$-NLS1s and 
FSRQs seems to suggest that similar processes (i.e., external-Compton dominated 
$\gamma$-ray emission) may be at work in both classes of sources 
\citep[e.g.,][]{Abdo2009:J0948discov,Berton2016:css_fsrlnls1,Arrieta2017}. 
Currently, seven FSRQs have been detected by IACTs 
(\citealt[][]{Mirzoyan:2017ton0599,Mukherjee2017:2017ton0599,Cerruti2017:arXiv170800658C,Ahnen2015:pks1441,VERITAS2015:pks1441,Sitarek2015:0218+35,MAGIC2011:pks1221+21,HESS2013:pks1510-089,Albert2008:3C279}; 
also see TeVCat\footnote{\href{http://tevcat.uchicago.edu/}{http://tevcat.uchicago.edu/}. } for further references). 
A significant detection of $\gamma$-NLS1 by CTA at $E \gtrsim 100$~GeV would then lead to 
similar conclusions about the location of the $\gamma$-ray emission region, i.e., near
the outer edge of the BLR.

\begin{figure*} 
\vspace{-4.7truecm}
\centerline{
 \hspace{+0.5truecm} 
                \includegraphics*[angle=0,width=20.5cm]{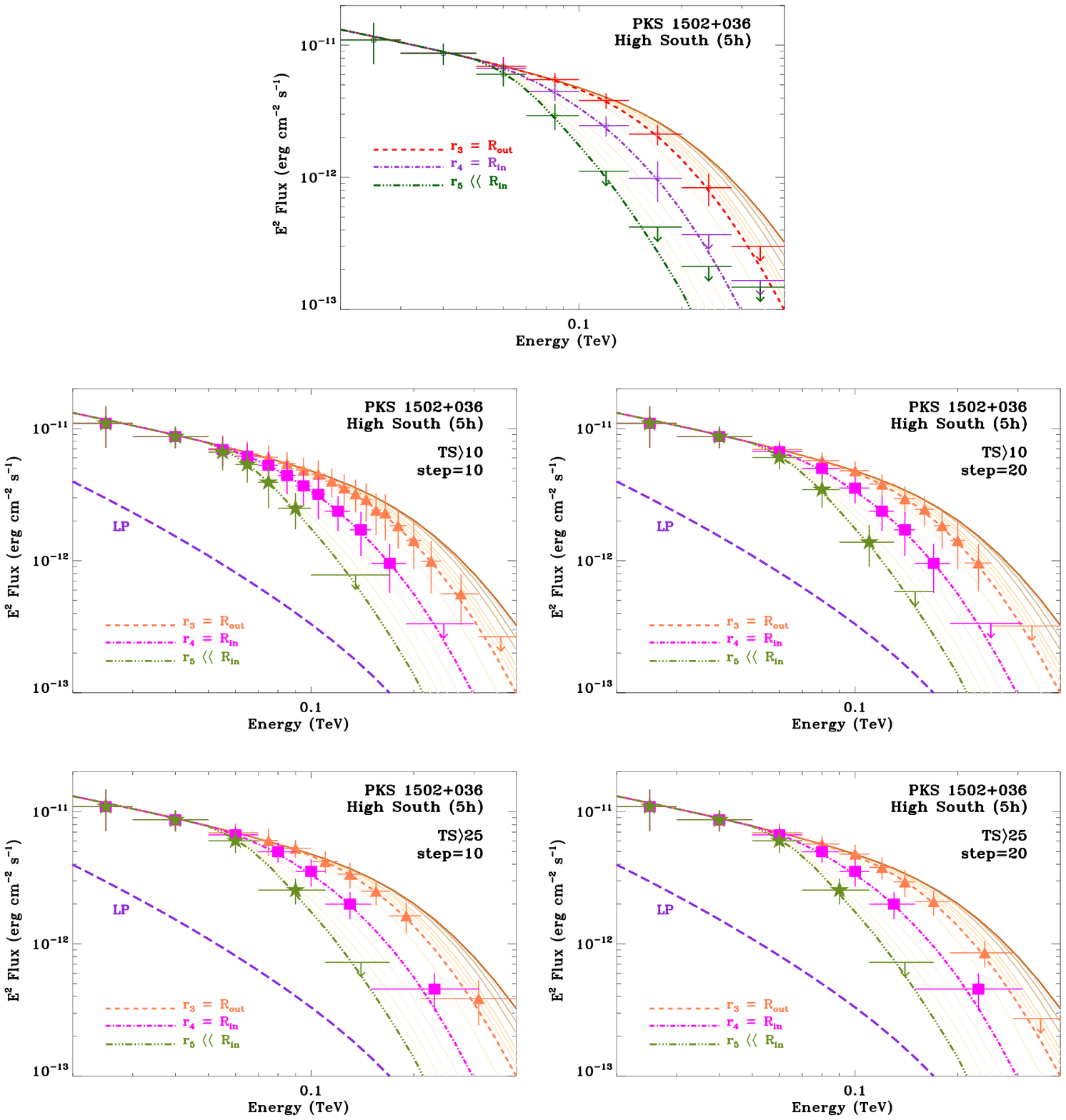} 
}
\vspace{-5truecm}
\caption[]{
Example of dynamical binning. \pks{} in the high state. 
The top panel is a zoom of Fig.~\ref{romano_nls1_2:fig:res_pks} (points in panels c,d, and e). 
The middle panels were obtained with dynamic energy binning by ensuring a $TS>10$ in each bin
and an energy step of 10\,GeV (left) and 20\,GeV (right). 
The bottom panels were obtained with dynamic energy binning by ensuring a $TS>25$ in each bin
and an energy step of 10\,GeV (left) and 20\,GeV (right). 
The violet long-dashed curve is the extrapolation of the 4FGL log-parabola fits scaled at the same Fermi flux. 
}
\label{romano_nls1_2:fig:res_dynbin} 
\end{figure*}

Looking at future perspectives, when the CTA data will be available, 
we sought to obtain a finer spectral binning by optimising the energy resolution with a 
dynamical binning to further improve our chances of disentangling the competing models.  
The method is described in detail in Appendix~\ref{romano_nls1_2:appendix_dynbin}.
Our results, obtained for the test case of \pks{} are shown in Fig.~\ref{romano_nls1_2:fig:res_dynbin}.
As a comparison with a static energy binning (Sect.~\ref{romano_nls1_2:simulations}), 
the top panel of Fig.~\ref{romano_nls1_2:fig:res_dynbin} 
summarises and zooms in on the points in panels c,d, and e of Fig.~\ref{romano_nls1_2:fig:res_pks}. 
The middle panels were obtained with dynamic energy binning by ensuring a $TS>10$ in each bin
and an energy step of 10\,GeV (left) and 20\,GeV (right). 
Similarly, the bottom panels were obtained with dynamic energy binning ensuring $TS>25$ in each bin. 
This method will clearly improve our chances of discriminating between 
competing models while simultaneously providing a better chance to study 
in detail the curvature of the spectrum.

In conclusion, we also note that an important contribution in our ability to 
distinguish among competing models can be obtained by a carefully planned, 
and strictly simultaneous, multi-wavelength campaign around high or flaring states in NLS1s. 
Indeed, the possible degeneracy within the VHE band can be lifted when optical--X-ray observations
are used to constrain the electron distribution based on the shape of the synchrotron portion of the spectrum.

        \section*{Acknowledgements}

%
We thank  J.\ Kn{\"o}dlseder and R.\ Terrier for helpful discussions, 
and  M.\ Cerruti, S.\ Razzaque, and J.\ Biteau  as internal CTA reviewer.  
PR and SV thank the staff at the Observatoire de Paris in Meudon, where part of the work was carried out,
and Amos, for keeping them on track. 
We acknowledge financial contribution from the agreement ASI-INAF n.\ 2017-14-H.0.  
This research has made use of the NASA/IPAC Extragalactic Database (NED) which 
           is operated by the Jet Propulsion Laboratory, California Institute of Technology, 
           under contract with the National Aeronautics and Space Administration. \\ 
This research made use of {\tt ctools}, a community-developed analysis 
           package for Imaging Air Cherenkov Telescope data. 
           ctools is based on {\tt GammaLib}, a community-developed toolbox for 
           the high-level analysis of astronomical gamma-ray data. \\ 
This research has made use of the CTA instrument response functions 
           provided by the CTA Consortium and Observatory, 
           see \href{https://www.cta-observatory.org/science/cta-performance/}{https://www.cta-observatory.org/science/cta-performance/} 
           (version prod3b-v1) for more details. \\ 
        We gratefully acknowledge financial support from the agencies 
          and organizations listed here: 
\href{http://www.cta-observatory.org/consortium_acknowledgments}{http://www.cta-observatory.org/consortium\_acknowledgments}.
This paper went through internal review by the CTA Consortium.\\ 
We also thank the anonymous referee for comments that helped improve the paper.


\bibliographystyle{mnras}

 \setcounter{table}{3} 
 \begin{table} 
\small
 \tabcolsep 2pt  
 \begin{center} 
 \caption{
Results of the simulations for the high state of \sbs. 
  \label{romano_nls1_2:table:results1}} 
  \begin{tabular}{lcccrr ccc c l } 
 \hline 
 \noalign{\smallskip} 
   Model                  &                Energy &  Det.\ c.l.$^{\mathrm{a}}$    & Det.\ c.l.$^{\mathrm{a}}$ & $\overline{TS_{\rm sim}}$  &E$^2$Flux$^{\mathrm{b}}$                \\   
                                                         & Range   &  (TS>10)                            & (TS>25)                         & & $\times10^{-13}$     \\ 
                                                         & (GeV)    & (\%)                                   &   (\%)                              &            &(erg\,cm$^{-2}$\,s$^{-1}$)       \\
 \noalign{\smallskip} 
 \hline 
 \noalign{\smallskip} 
r1
 &     20--30 &  100.0 &   91.0 &  $    41.2\pm 11.5 $  &   $     89.6\pm    14.5 $  \\
 &     30--50 &  100.0 &  100.0 &  $   126.3\pm 21.7 $  &   $     82.4\pm     7.7 $  \\
 &     50--70 &  100.0 &  100.0 &  $   224.0\pm 26.9 $  &   $     74.0\pm     4.8 $  \\
 &     70--100 &  100.0 &  100.0 &  $   433.4\pm 38.9 $  &   $     64.7\pm     3.1 $  \\
 &    100--140 &  100.0 &  100.0 &  $   587.1\pm 44.6 $  &   $     50.8\pm     2.0 $  \\
 &    140--200 &  100.0 &  100.0 &  $   624.6\pm 54.6 $  &   $     32.3\pm     1.5 $  \\
 &    200--280 &  100.0 &  100.0 &  $   264.9\pm 34.3 $  &   $     14.9\pm     1.1 $  \\
 &    280--400 &  100.0 &   99.0 &  $    59.8\pm 15.9 $  &   $      4.6\pm     0.8 $  \\
  \noalign{\smallskip} 
r2
 &     50--70 &  100.0 &  100.0 &  $   224.0\pm 26.9 $  &   $     74.0\pm     4.8 $  \\
 &     70--100 &  100.0 &  100.0 &  $   433.4\pm 38.9 $  &   $     64.7\pm     3.1 $  \\
 &    100--140 &  100.0 &  100.0 &  $   587.1\pm 44.6 $  &   $     50.8\pm     2.0 $  \\
 &    140--200 &  100.0 &  100.0 &  $   624.6\pm 54.6 $  &   $     32.3\pm     1.5 $  \\
 &    200--280 &  100.0 &  100.0 &  $   264.9\pm 34.3 $  &   $     14.9\pm     1.1 $  \\
 &    280--400 &  100.0 &   99.0 &  $    59.5\pm 16.3 $  &   $      4.6\pm     0.8 $  \\
 \noalign{\smallskip} 
 r3
 &     50--70 &  100.0 &  100.0 &  $   223.5\pm 27.0 $  &   $     73.9\pm     4.8 $  \\
 &     70--100 &  100.0 &  100.0 &  $   417.6\pm 36.5 $  &   $     63.6\pm     3.0 $  \\
 &    100--140 &  100.0 &  100.0 &  $   457.5\pm 39.2 $  &   $     44.5\pm     2.0 $  \\
 &    140--200 &  100.0 &  100.0 &  $   292.7\pm 38.9 $  &   $     20.9\pm     1.5 $  \\
 &    200--280 &  100.0 &   98.0 &  $    51.6\pm 14.3 $  &   $      5.7\pm     1.0 $  \\
 &    280--400 &   14.8 &    0.0 &  $     5.1\pm  3.9 $  &   $      <1.1$  \\
  \noalign{\smallskip} 
r4
 &     50--70 &  100.0 &  100.0 &  $   193.2\pm 24.6 $  &   $     69.2\pm     4.7 $  \\
 &     70--100 &  100.0 &  100.0 &  $   215.1\pm 27.4 $  &   $     45.0\pm     2.9 $  \\
 &    100--140 &  100.0 &  100.0 &  $   107.4\pm 20.3 $  &   $     20.4\pm     2.1 $  \\
 &    140--200 &   99.0 &   62.0 &  $    30.3\pm 11.4 $  &   $      5.9\pm     1.3 $  \\
 &    200--280 &    1.3 &    0.0 &  $     3.2\pm  2.8 $  &   $      <1.1  $  \\
 &    280--400 &    0.0 &    0.0 &  $     1.8\pm  1.8 $  &   $      <0.7 $  \\
  \noalign{\smallskip} 
r5
 &     50--70 &  100.0 &  100.0 &  $   141.6\pm 21.5 $  &   $     59.4\pm     4.7 $  \\
 &     70--100 &  100.0 &  100.0 &  $    60.4\pm 14.5 $  &   $     21.9\pm     3.0 $  \\
 &    100--140 &   31.0 &    1.0 &  $     8.4\pm  5.4 $  &   $      <5.0 $  \\
 &    140--200 &    0.0 &    0.0 &  $     2.6\pm  2.3 $  &   $      <1.5  $  \\
 &    200--280 &    0.0 &    0.0 &  $     1.5\pm  1.4 $  &   $      <0.8  $  \\
 &    280--400 &    0.0 &    0.0 &  $     1.7\pm  1.7 $  &   $      <0.6  $  \\

 \noalign{\smallskip}
  \hline
  \end{tabular}
\end{center}
\begin{list}{\it Notes.}{} 
  \item[$^{\mathrm{a}}$]{We consider a detection to have a high significance when $TS \geq 25$  
and a low significance when $10 \leq TS < 25$. 
The source will not be considered detected for $TS <10$. }
  \item[$^{\mathrm{b}}$]{Upper limits are calculated for 95\,\% confidence level for all cases where $TS <10$.  }
 \end{list}   
  \end{table}                
%
\setcounter{table}{4} 
 \begin{table} 
\small
 \tabcolsep 2pt  
 \begin{center} 
 \caption{
Results of the simulations for the flare state of \pmn. 
  \label{romano_nls1_2:table:results2}} 
  \begin{tabular}{lcccrr ccc c l } 
 \hline 
 \noalign{\smallskip} 
   Model                  &                Energy &  Det.\ c.l.$^{\mathrm{a}}$    & Det.\ c.l.$^{\mathrm{a}}$ & $\overline{TS_{\rm sim}}$  &E$^2$Flux$^{\mathrm{b}}$                \\   
                                                         & Range   &  (TS>10)                            & (TS>25)                         & & $\times10^{-13}$     \\ 
                                                         & (GeV)    & (\%)                                   &   (\%)                              &            &(erg\,cm$^{-2}$\,s$^{-1}$)       \\
 \noalign{\smallskip} 
 \hline 
 \noalign{\smallskip} 
r1
 &     20--30 &  100.0 &  100.0 &  $    66.2\pm 17.2 $  &   $    361.8\pm    49.4 $  \\
 &     30--50 &  100.0 &  100.0 &  $   169.8\pm 27.3 $  &   $    275.0\pm    23.2 $  \\
 &     50--70 &  100.0 &  100.0 &  $   201.7\pm 29.9 $  &   $    207.3\pm    16.7 $  \\
 &     70--100 &  100.0 &  100.0 &  $   282.1\pm 37.6 $  &   $    155.9\pm    11.7 $  \\
 &    100--140 &  100.0 &  100.0 &  $   274.9\pm 36.9 $  &   $    106.4\pm     8.3 $  \\
 &    140--200 &  100.0 &  100.0 &  $   198.0\pm 33.1 $  &   $     57.9\pm     5.4 $  \\
 &    200--280 &  100.0 &  100.0 &  $    74.8\pm 21.3 $  &   $     23.4\pm     4.1 $  \\
 &    280--400 &   71.7 &   15.2 &  $    15.8\pm  8.8 $  &   $      6.6\pm     2.4 $  \\
\noalign{\smallskip} 
 r2
 &     50--70 &  100.0 &  100.0 &  $   201.7\pm 29.9 $  &   $    207.3\pm    16.7 $  \\
 &     70--100 &  100.0 &  100.0 &  $   282.1\pm 37.6 $  &   $    155.9\pm    11.7 $  \\
 &    100--140 &  100.0 &  100.0 &  $   274.9\pm 36.9 $  &   $    106.4\pm     8.3 $  \\
 &    140--200 &  100.0 &  100.0 &  $   198.0\pm 33.1 $  &   $     57.9\pm     5.4 $  \\
 &    200--280 &  100.0 &  100.0 &  $    74.8\pm 21.3 $  &   $     23.4\pm     4.1 $  \\
 &    280--400 &   70.7 &   13.1 &  $    15.6\pm  8.8 $  &   $      6.5\pm     2.4 $  \\
 \noalign{\smallskip} 
r3
 &     50--70 &  100.0 &  100.0 &  $   196.3\pm 30.4 $  &   $    204.5\pm    16.9 $  \\
 &     70--100 &  100.0 &  100.0 &  $   231.9\pm 34.9 $  &   $    140.0\pm    11.9 $  \\
 &    100--140 &  100.0 &  100.0 &  $   124.0\pm 23.6 $  &   $     66.4\pm     7.1 $  \\
 &    140--200 &   94.0 &   60.0 &  $    27.5\pm 11.4 $  &   $     16.9\pm     4.6 $  \\
 &    200--280 &    1.3 &    0.0 &  $     3.1\pm  2.7 $  &   $      <3.4$  \\
 &    280--400 &    0.0 &    0.0 &  $     1.8\pm  1.7 $  &   $      <1.8$  \\
 \noalign{\smallskip} 
r4
 &     50--70 &  100.0 &  100.0 &  $   164.3\pm 23.4 $  &   $    186.4\pm    14.2 $  \\
 &     70--100 &  100.0 &  100.0 &  $   104.7\pm 22.7 $  &   $     87.8\pm    10.5 $  \\
 &    100--140 &   92.0 &   39.0 &  $    22.8\pm  9.6 $  &   $     24.0\pm     6.7 $  \\
 &    140--200 &    4.9 &    0.0 &  $     3.6\pm  2.8 $  &   $      <5.4$  \\
 &    200--280 &    0.0 &    0.0 &  $     1.9\pm  1.9 $  &   $      <2.7$  \\
 &    280--400 &    0.0 &    0.0 &  $     1.7\pm  1.7 $  &   $      <1.8$  \\
 \noalign{\smallskip} 
r5
 &     50--70 &  100.0 &  100.0 &  $   107.5\pm 21.2 $  &   $    145.5\pm    15.8 $  \\
 &     70--100 &   78.0 &   18.0 &  $    17.2\pm  8.9 $  &   $     28.5\pm     9.6 $  \\
 &    100--140 &    0.0 &    0.0 &  $     2.1\pm  2.1 $  &   $      <6.1$  \\
 &    140--200 &    0.0 &    0.0 &  $     1.5\pm  1.9 $  &   $      <3.2$  \\
 &    200--280 &    0.0 &    0.0 &  $     2.0\pm  1.9 $  &   $      <2.7$  \\
 &    280--400 &    0.0 &    0.0 &  $     1.7\pm  1.7 $  &   $      <1.7$  \\
 \noalign{\smallskip}
  \hline
  \end{tabular}
\end{center}
\begin{list}{\it Notes.}{} 
  \item[$^{\mathrm{a}}$]{We consider a detection to have a high significance when $TS \geq 25$  
and a low significance when $10 \leq TS < 25$. 
The source will not be considered detected for $TS <10$. }
  \item[$^{\mathrm{b}}$]{Upper limits are calculated for 95\,\% confidence level for all cases where $TS <10$.  }
 \end{list}   
  \end{table}                
%
\setcounter{table}{5} 
 \begin{table} 
\small
 \tabcolsep 2pt  
 \begin{center} 
 \caption{
Results of the simulations for the high state of \pmn. 
  \label{romano_nls1_2:table:results3}} 
  \begin{tabular}{lcccrr ccc c l } 
 \hline 
 \noalign{\smallskip} 
   Model                  &                Energy &  Det.\ c.l.$^{\mathrm{a}}$    & Det.\ c.l.$^{\mathrm{a}}$ & $\overline{TS_{\rm sim}}$  &E$^2$Flux$^{\mathrm{b}}$                \\   
                                                         & Range   &  (TS>10)                            & (TS>25)                         & & $\times10^{-13}$     \\ 
                                                         & (GeV)    & (\%)                                   &   (\%)                              &            &(erg\,cm$^{-2}$\,s$^{-1}$)       \\
 \noalign{\smallskip} 
 \hline 
 \noalign{\smallskip} 
r1
 &     20--30 &   69.0 &    9.0 &  $    14.0\pm  7.4 $  &   $    114.3\pm    37.2 $  \\
 &     30--50 &  100.0 &   79.0 &  $    33.9\pm 10.9 $  &   $     88.6\pm    15.9 $  \\
 &     50--70 &  100.0 &   94.0 &  $    43.5\pm 13.5 $  &   $     68.8\pm    12.0 $  \\
 &     70--100 &  100.0 &  100.0 &  $    63.8\pm 16.0 $  &   $     52.2\pm     7.2 $  \\
 &    100--140 &  100.0 &  100.0 &  $    65.2\pm 15.5 $  &   $     35.9\pm     4.9 $  \\
 &    140--200 &  100.0 &   96.0 &  $    50.9\pm 16.4 $  &   $     19.9\pm     3.6 $  \\
 &    200--280 &   80.0 &   23.0 &  $    18.8\pm  9.5 $  &   $      7.8\pm     2.5 $  \\
 &    280--400 &   16.7 &    0.0 &  $     5.6\pm  4.7 $  &   $      <2.9 $  \\
 \noalign{\smallskip} 
r2
 &     50--70 &  100.0 &   94.0 &  $    43.5\pm 13.5 $  &   $     68.8\pm    12.0 $  \\
 &     70--100 &  100.0 &  100.0 &  $    63.8\pm 16.0 $  &   $     52.2\pm     7.2 $  \\
 &    100--140 &  100.0 &  100.0 &  $    65.2\pm 15.5 $  &   $     35.9\pm     4.9 $  \\
 &    140--200 &  100.0 &   96.0 &  $    50.9\pm 16.4 $  &   $     19.9\pm     3.6 $  \\
 &    200--280 &   80.0 &   23.0 &  $    18.8\pm  9.5 $  &   $      7.8\pm     2.5 $  \\
 &    280--400 &   15.6 &    0.0 &  $     5.6\pm  4.6 $  &   $      <2.9 $  \\
 \noalign{\smallskip} 
r3
 &     50--70 &  100.0 &   93.0 &  $    42.5\pm 12.7 $  &   $     68.0\pm    11.4 $  \\
 &     70--100 &  100.0 &   98.0 &  $    51.5\pm 13.1 $  &   $     46.5\pm     6.8 $  \\
 &    100--140 &   99.0 &   61.0 &  $    29.3\pm 10.2 $  &   $     22.6\pm     4.7 $  \\
 &    140--200 &   27.8 &    0.0 &  $     7.3\pm  5.5 $  &   $      <6.4 $  \\
 &    200--280 &    1.8 &    0.0 &  $     2.2\pm  2.2 $  &   $      <2.2 $  \\
 &    280--400 &    0.0 &    0.0 &  $     1.8\pm  1.7 $  &   $      <1.4 $  \\
\noalign{\smallskip} 
r4
 &     50--70 &  100.0 &   83.0 &  $    35.3\pm 11.1 $  &   $     61.3\pm    11.2 $  \\
 &     70--100 &   96.0 &   38.0 &  $    22.5\pm  8.1 $  &   $     28.8\pm     6.5 $  \\
 &    100--140 &   20.4 &    0.0 &  $     6.2\pm  4.4 $  &   $      <9.4 $  \\
 &    140--200 &    1.6 &    0.0 &  $     2.6\pm  2.3 $  &   $      <3.3 $  \\
 &    200--280 &    2.0 &    0.0 &  $     1.7\pm  2.1 $  &   $      <1.8 $  \\
 &    280--400 &    0.0 &    0.0 &  $     1.7\pm  1.7 $  &   $      <1.4 $  \\
\noalign{\smallskip} 
 r5
 &     50--70 &   96.0 &   43.0 &  $    23.2\pm  8.9 $  &   $     47.9\pm    10.5 $  \\
 &     70--100 &    7.4 &    0.0 &  $     4.4\pm  3.4 $  &   $     <11.3 $  \\
 &    100--140 &    0.0 &    0.0 &  $     1.7\pm  1.5 $  &   $      <4.4 $  \\
 &    140--200 &    0.0 &    0.0 &  $     1.9\pm  1.9 $  &   $      <2.6 $  \\
 &    200--280 &    2.1 &    0.0 &  $     1.7\pm  2.1 $  &   $      <1.8 $  \\
 &    280--400 &    0.0 &    0.0 &  $     1.8\pm  1.7 $  &   $      <1.4 $  \\
 \noalign{\smallskip}
  \hline
  \end{tabular}
\end{center}
\begin{list}{\it Notes.}{} 
  \item[$^{\mathrm{a}}$]{We consider a detection to have a high significance when $TS \geq 25$  
and a low significance when $10 \leq TS < 25$. 
The source will not be considered detected for $TS <10$. }
  \item[$^{\mathrm{b}}$]{Upper limits are calculated for 95\,\% confidence level for all cases where $TS <10$.  }
 \end{list}   
  \end{table}                
%
\setcounter{table}{6} 
 \begin{table} 
\small
 \tabcolsep 2pt  
 \begin{center} 
 \caption{
Results of the simulations for the high state of \pks. 
  \label{romano_nls1_2:table:results4}} 
  \begin{tabular}{lcccrr ccc c l } 
 \hline 
 \noalign{\smallskip} 
   Model                  &                Energy &  Det.\ c.l.$^{\mathrm{a}}$    & Det.\ c.l.$^{\mathrm{a}}$ & $\overline{TS_{\rm sim}}$  &E$^2$Flux$^{\mathrm{b}}$                \\   
                                                         & Range   &  (TS>10)                            & (TS>25)                         & & $\times10^{-13}$     \\ 
                                                         & (GeV)    & (\%)                                   &   (\%)                              &            &(erg\,cm$^{-2}$\,s$^{-1}$)       \\
 \noalign{\smallskip} 
 \hline 
 \noalign{\smallskip} 
r1
 &     20--30 &   64.0 &   10.0 &  $    13.2\pm  7.1 $  &   $    109.4\pm    37.7 $  \\
 &     30--50 &  100.0 &   80.0 &  $    32.7\pm 10.5 $  &   $     86.9\pm    15.8 $  \\
 &     50--70 &  100.0 &   95.0 &  $    44.6\pm 13.8 $  &   $     69.3\pm    12.0 $  \\
 &     70--100 &  100.0 &  100.0 &  $    70.7\pm 16.4 $  &   $     55.0\pm     7.0 $  \\
 &    100--140 &  100.0 &  100.0 &  $    84.1\pm 17.7 $  &   $     41.3\pm     5.0 $  \\
 &    140--200 &  100.0 &  100.0 &  $    84.3\pm 21.2 $  &   $     26.8\pm     3.6 $  \\
 &    200--280 &  100.0 &   95.0 &  $    53.1\pm 16.6 $  &   $     14.6\pm     2.7 $  \\
 &    280--400 &   83.0 &   33.0 &  $    20.0\pm 10.0 $  &   $      5.9\pm     1.9 $  \\
\noalign{\smallskip} 
 r2
 &     50--70 &  100.0 &   95.0 &  $    44.6\pm 13.8 $  &   $     69.3\pm    12.0 $  \\
 &     70--100 &  100.0 &  100.0 &  $    70.7\pm 16.4 $  &   $     55.0\pm     7.0 $  \\
 &    100--140 &  100.0 &  100.0 &  $    84.1\pm 17.7 $  &   $     41.3\pm     5.0 $  \\
 &    140--200 &  100.0 &  100.0 &  $    84.3\pm 21.2 $  &   $     26.8\pm     3.6 $  \\
 &    200--280 &  100.0 &   95.0 &  $    53.1\pm 16.6 $  &   $     14.6\pm     2.7 $  \\
 &    280--400 &   83.0 &   33.0 &  $    20.0\pm 10.0 $  &   $      5.9\pm     1.9 $  \\
\noalign{\smallskip} 
 r3
 &     50--70 &  100.0 &   95.0 &  $    44.4\pm 13.7 $  &   $     69.2\pm    11.9 $  \\
 &     70--100 &  100.0 &  100.0 &  $    70.9\pm 16.2 $  &   $     55.0\pm     6.9 $  \\
 &    100--140 &  100.0 &  100.0 &  $    72.7\pm 17.2 $  &   $     38.2\pm     5.1 $  \\
 &    140--200 &  100.0 &   98.0 &  $    57.0\pm 17.6 $  &   $     21.2\pm     3.7 $  \\
 &    200--280 &   86.0 &   35.0 &  $    20.9\pm  9.1 $  &   $      8.4\pm     2.3 $  \\
 &    280--400 &   19.3 &    0.0 &  $     6.0\pm  4.5 $  &   $      <3.0 $  \\
 \noalign{\smallskip} 
r4
 &     50--70 &  100.0 &   94.0 &  $    41.3\pm 11.6 $  &   $     66.9\pm    10.9 $  \\
 &     70--100 &  100.0 &   97.0 &  $    47.5\pm 12.6 $  &   $     44.6\pm     6.6 $  \\
 &    100--140 &  100.0 &   81.0 &  $    33.2\pm  9.9 $  &   $     24.7\pm     4.3 $  \\
 &    140--200 &   75.0 &   15.0 &  $    15.9\pm  8.4 $  &   $      9.8\pm     3.3 $  \\
 &    200--280 &   13.8 &    0.0 &  $     5.1\pm  4.0 $  &   $      <3.7 $  \\
 &    280--400 &    0.0 &    0.0 &  $     2.4\pm  2.3 $  &   $      <1.7 $  \\
 \noalign{\smallskip} 
r5
 &     50--70 &  100.0 &   80.0 &  $    34.2\pm 11.2 $  &   $     60.3\pm    11.4 $  \\
 &     70--100 &   98.0 &   40.0 &  $    23.1\pm  8.3 $  &   $     29.3\pm     6.5 $  \\
 &    100--140 &   31.3 &    0.0 &  $     7.9\pm  4.9 $  &   $     <11.1  $  \\
 &    140--200 &    6.2 &    0.0 &  $     3.6\pm  3.3 $  &   $      <4.2 $  \\
 &    200--280 &    1.8 &    0.0 &  $     2.2\pm  2.2 $  &   $      <2.1 $  \\
 &    280--400 &    0.0 &    0.0 &  $     1.8\pm  1.7 $  &   $      <1.5 $  \\
 \noalign{\smallskip}
  \hline
  \end{tabular}
\end{center}
\begin{list}{\it Notes.}{} 
  \item[$^{\mathrm{a}}$]{We consider a detection to have a high significance when $TS \geq 25$  
and a low significance when $10 \leq TS < 25$. 
The source will not be considered detected for $TS <10$. }
  \item[$^{\mathrm{b}}$]{Upper limits are calculated for 95\,\% confidence level for all cases where $TS <10$.  }
 \end{list}   
  \end{table}                
%

\clearpage
\appendix

 	 \section{Finer Spectral characterisation with optimisation of the energy bins via dynamical binning} 
                                \label{romano_nls1_2:appendix_dynbin}

In the following we consider the case of \pks{} in high state to propose an iterative 
method to optimise the choice of the energy bins for spectrum calculations.
While this may seem a purely academic pursuit at this point, it may yield 
the necessary edge to better discriminate among competing models
when the data from CTA will be available. 
Furthermore, adaptive binning is commonly used in high energy astrophysics 
\citep[see][for {\it Fermi} light curves, rather than spectra, as we did in this paper]{Lott2012:adaptive_bin}.

We start from the lowest energy boundary, $E_{0}$,
and select a small increment in energy $dE$ (or 'step'). We then perform a `detection', i.e., we 
\begin{enumerate}
\item create an event list with  {\tt ctobssim} with energies between $E_0$ and $E_0+dE$,  [$E_0$,$E_0+dE$]; 
\item perform a maximum likelihood model fitting with {\tt ctlike}; 
\item check if the source is detected above a certain TS  threshold value, $TS \ga TS_{\rm thresh}$. 
        \begin{enumerate}
        \item  If the source is detected ($TS\ga TS_{\rm thresh}$), then we repeat for the next bin, 
                   defined by a lower energy boundary $E_0+dE$ 
                   and the same energy increment, [$E_0+dE$, $E_0+2\,dE$]. 
        \item  if the source is not detected we repeat in the energy bin  [$E_0$,$E_0+2\,dE$], 
                   or [$E_0$,$E_0+n\,dE$], with $n$ large enough that $TS\ga TS_{\rm thresh}$. 
        \end{enumerate}
\end{enumerate}
The procedure is repeated until the high energy boundary of the bin reaches 
a reasonable $E_{\rm max}$, as defined by the highest detection obtained without a dynamical binning. 

For our example (Fig.~\ref{romano_nls1_2:fig:res_dynbin}), we used the models $r_3 =R_{\rm out}$, $r_4 =R_{\rm in}$, and  $r_5\ll R_{\rm in}$ as input models. 
We selected $E_0=50$\,GeV, since below this energy the models are virtually indistinguishable, 
and $E_{\rm max}=400$\,GeV.  
Also, as a compromise between speed of processing and achieving a fine resolution in the resulting spectrum, 
we adopted a step of $dE=10$\,GeV and $dE=20$\,GeV. 
For the last bin, we calculated the 95\,\% confidence level upper limits on fluxes 
from the distribution of the simulated fluxes (see Sect.~\ref{romano_nls1_2:simulations}). 
We considered both cases of a high significance detection ($TS_{\rm thresh}=25$) and the
low significance detection ($TS_{\rm thresh}=10$).

\bsp	
\label{lastpage}
\end{document}